\documentclass[useAMS,usegraphicx,usenatbib]{mn2e}

\usepackage{url}
\usepackage{times}

\usepackage{graphicx}

\usepackage{amssymb,amsmath}

\def\refjnl#1{{\rm#1}}%
\newcommand\araa{\refjnl{ARA\&A}}%
\newcommand\apj{\refjnl{ApJ}}%
\newcommand\apjl{\refjnl{ApJ}}%
\newcommand\apjs{\refjnl{ApJS}}%
\newcommand\apss{\refjnl{Ap\&SS}}%
\newcommand\aap{\refjnl{A\&A}}%
\newcommand\aapr{\refjnl{A\&A~Rev.}}%
\newcommand\mnras{\refjnl{MNRAS}}%
\newcommand\prd{\refjnl{Phys.~Rev.~D}}%
\newcommand\nat{\refjnl{Nature}}%
\newcommand\jcap{\refjnl{J.~Cosmology~Astropart.~Phys.}}%

\def\beq{\begin{equation}}
\def\eeq{\end{equation}}

\newcommand{\tableline}{\hline}
\newcommand{\boldtext}{}

\title{Detecting Dark Matter Substructures around the Milky Way with Gaia} 

\author[R. Feldmann and D. Spolyar]{Robert Feldmann,$^{1,\ast}$\thanks{Hubble fellow} and Douglas Spolyar$^{2,\ast}$
\\
\normalsize{$^{1}$Department of Astronomy, University of California, Berkeley, CA 94720-3411, USA}\\
\normalsize{$^{2}$Institut d'astrophysique de Paris, Paris, 75014, France}\\
\\
\normalsize{$^\ast$Corresponding author. E-mail: feldmann@berkeley.edu (R.F.); dspolyar@iap.fr (D.S.)}
}

\begin{document} 

\maketitle 

\begin{abstract}
Cold Dark Matter (CDM) theory, a pillar of modern cosmology and astrophysics, predicts the existence of a large number of starless dark matter halos surrounding the Milky Way (MW). However, clear observational evidence of these ``dark'' substructures remains elusive. Here, we present a detection method based on the small, but detectable, velocity changes that an orbiting substructure imposes on the stars in the MW disk. Using high-resolution numerical simulations we estimate that the new space telescope Gaia should detect the kinematic signatures of a few starless substructures provided the CDM paradigm holds. Such a measurement will provide unprecedented constraints on the primordial matter power spectrum at low-mass scales and offer a new handle onto the particle physics properties of dark matter.
\end{abstract}


\section{Introduction}
Substructures and satellite galaxies interacting and merging with the MW can significantly affect the dynamical state of its stellar disk (e.g., \citealt{1993ApJ...403...74Q}). Such gravitational interactions may result in tidal heating of the disk \citep{1985ApJ...299..633L, 1987ApJ...316...23C, 1992ApJ...389....5T, 1993ApJ...403...74Q, 1996ApJ...460..121W, 2004MNRAS.351.1215B}, in the excitation of bending waves (e.g., \citealt{1998ApJ...506..590S}), in tilts and warps (e.g., \citealt{1997ApJ...480..503H}), in flaring (e.g., \citealt{2008ApJ...688..254K, 2008ASPC..396..321D}), or may trigger the growth of non-axisymmetric structures such as bars \citep{1996ApJ...460..121W, 2002ApJ...574L..43C, 2006ApJ...653.1180G, 2008ApJ...688..254K} and ring-like stellar enhancements (e.g., \citealt{2008ApJ...688..254K, 2011Natur.477..301P}). 

Recent high resolution N-body simulations have also shown that massive substructures colliding with the MW disk may excite wave-like changes of the stellar density and velocity components of disk stars \citep{2008ASPC..396..321D, 2011Natur.477..301P, 2012MNRAS.419.2163G}. Specifically, it has been suggested that the passage of the Sagittarius dwarf spheroidal galaxy ($M_{\rm vir}\lesssim10^{11}$ $M_\odot$ at infall into the MW halo;  \citealt{2000MNRAS.314..468J}) may be responsible for the ringing of the MW disk \citep{2011Natur.477..301P, 2012MNRAS.423.3727G}. Gravitational encounters with such comparably massive satellite galaxies are also suspected to be the cause of the observed north/south asymmetries in the stellar number densities and mean vertical and radial stellar velocities \citep{2012ApJ...750L..41W, 2013MNRAS.429..159G, 2013MNRAS.436..101W, 2013ApJ...777L...5C, 2013ApJ...777...91Y}.

While massive satellite galaxies have the largest impact on the dynamical state of the MW disk, they are relatively rare at the present epoch \citep{2008ApJ...688..254K}. Low mass substructures are more numerous and they have a qualitatively similar (although significantly weaker) effect on stars in the MW disk. Low mass substructures ($M_{\rm vir}\lesssim{}10^{9}$ $M_\odot$) are hard to detect in the electromagnetic spectrum because they are largely devoid of gas and stars as a result of the increase in the thermal Jeans mass following re-ionization \citep{1999ApJ...523...54B, 2008MNRAS.390..920O}. Proposed methods to detect such substructures around the Milky Way (MW) rely on gamma-ray emission from the annihilation of dark matter \citep{1990Natur.346...39L, 2000PhRvD..62l3005C, 2007ApJ...657..262D}, or on the gravitational scattering of stars in the tidal streams of satellite galaxies \citep{2002MNRAS.332..915I,2002ApJ...570..656J,2008ApJ...681...40S}.

The unknown nature of dark matter and its annihilation channels, as well as the large background of gamma-rays from more conventional astrophysical sources, poses major challenges for the former approach \citep{2012JCAP...11..050Z}. In contrast, the low signal-to-noise ratio of the available observational data and the low number of suitable streams limit the use of tidal streams to detect substructures \citep{2013ApJ...768..171C}. So far, neither approach has produced definite evidence in favor of truly starless dark matter halos orbiting the MW.

The observational difficulties are substantial, yet, the identification of starless, low-mass substructures will have profound implications for the understanding of dark matter. For instance, it will provide direct evidence for the existence of dark matter that is clustered on small scales. Furthermore, the number density of dark matter halos encodes invaluable information about the primordial power spectrum, the physics of the early universe, and the nature of dark matter \citep{1999ApJ...524L..19M,2000ApJ...539..517B, 2012PDU.....1...50K}. For instance, in Warm Dark Matter models, a competitor of the Cold Dark Matter (CDM) paradigm, structure formation is suppressed below the free-streaming scale of the dark matter particle, resulting in a deficit in substructure with masses below $\sim{}10^{9}$ $M_\odot$ \citep{2001ApJ...556...93B, 2003ApJ...598...49Z, 2005PhRvD..71f3534V}.

We propose to detect dark substructures as they pass through the disk of the MW based on their gravitational pull on disk stars. Specifically, we will show that such substructures leave a tell-tale kinematic imprint in the velocity field of disk stars that could potentially be measured with the large-scale, high-precision astrometric mission Gaia. The main idea is straightforward. An object that passes with relative speed $V$ through the stellar disk induces a localized velocity impulse in the disk of about (see Appendix \ref{app:VelChange})

\begin{flalign}
\Delta v_* & \sim{} \frac{2\,G\,M}{Vb}f \nonumber \\
                          & \simeq{} 2.2\,\mathrm{km\,s^{-1}} \left(\frac{M}{10^8\,M_\odot}\right)\left(\frac{1\,\mathrm{kpc}}{b}\right)\left(\frac{400\,\mathrm{km\,s^{-1}}}{V}\right)f.
\label{eq:DeltaV}
\end{flalign}

Here $G$ is the Newton constant, $f$ is a factor of order unity that depends on the orbit of the perturber, $b$ is the impact parameter of the encounter for a star in the disk, and $M$ is a characteristic mass of the passing object. For a point-like perturber, $M$ is the total mass. For an object with an extended but steeply radially declining density profile, $M$ is approximately the mass within $b$.

Velocity changes caused by passing low-mass substructures are thus smaller than the velocity dispersion of the stars in the disk ($\sim{}$25 km s$^{-1}$ in the solar neighborhood, e.g., \citealt{2013A&ARv..21...61R}). However, as stars in the same vicinity (within 1-2 kpc for a $10^8$ $M_\odot$ perturber) experience approximately the same force, the kinematic signature of the substructure may in principle be recovered by spatially averaging a sufficiently large sample of disk star velocities.

Measuring this kinematic imprint offers a variety of advantages compared with approaches that infer the presence of perturbers from the excitation of kinematic or density waves in the stellar disk. First, the velocity perturbations have a unique morphology that enables us to differentiate them from disturbances caused by, e.g., spiral structure or a stellar bar. Second, the kinematic signal localized (to within a few kpc) before winding sets in
and, hence, can be used to track where the substructure passed through the disk. Third, the signal has a life time of $\sim{}100$ Myr. This time is long enough to make it likely that we can observe the imprint of one or several substructures crossing the disk at any given time. It is also short enough to erase the memory of the multitude of previous encounters and, hence, presents the disk as a clean slate every $\sim{}$half dynamical time. In contrast, bending modes and density waves likely survive for several rotation periods (e.g. \citealt{1969ApJ...155..747H, 1969ApJ...158..899T, 1977ARA&A..15..437T, 1984ApJ...280..117S}). Furthermore, although the observed kinematic asymmetries in the radial and vertical directions (e.g., \citealt{2012ApJ...750L..41W}) could be caused by an external perturber, it is also possible that they are excited by internal non-axisymmetric features of the MW disk \citep{2014MNRAS.440.2564F}. The localized, short-lived kinematic velocity impulse that gives rise to equation (\ref{eq:DeltaV}) avoids these problems and thus minimizes the number of false positive detections of dark substructures.

In this paper we investigate the feasibility of the proposed detection method using high resolution numerical simulations and mock stellar catalogs.
We introduce the numerical set-up in section \ref{sect:NumSetup}. In section \ref{sect:CompSims} we analyze the kinematic signature that the passing substructure imparts on the stellar disk. We estimate the rate of substructure collisions with the MW in section \ref{sect:Collisions}. We discuss the implementation of the proposed method with a Gaia-based survey in section \ref{sect:Gaia}. We summarize our findings and conclude in section \ref{sect:Summary}.

\section{Numerical set-up}
\label{sect:NumSetup}

Our numerical set-up consists of models of $(i)$ a dynamically stable, dissipationless galaxy with properties similar to the MW \citep{2008ApJ...679.1239W} and $(ii)$ a dark matter substructure with virial mass $1.1\times{}10^{9}$ $M_\odot$, scale radius 1.3 kpc, mass within the scale radius of $1.1\times{}10^{8}$ $M_\odot$, and virial circular velocity 15 km s$^{-1}$, see Appendix \ref{app:ModelSetup}. The mass ($\sim{}10^4$ $M_\odot$) and force resolution ($\sim{}20-50$ pc) of our numerical set-up are adequate to follow accurately the dynamical evolution of the coupled MW -- substructure system. We summarize the resolution of the simulations in Table~\ref{tab:Res}. 

\begin{table}
\begin{center}
\begin{tabular}{c|c|c|c}
\tableline \tableline
component & total No. of particles & particle mass & softening \\ 
 &  & $M_\odot$ & pc \\ \tableline
MW disk &  $8\times{}10^6$ & $4.6\times{}10^3$  & 20 \\
MW bulge &  $2\times{}10^6$ & $5.1\times{}10^3$  & 20 \\
MW halo & $2\times{}10^7$  & $5.1\times{}10^4$ & 50 \\
substructure &  $1.2\times{}10^5$ & $1.0\times{}10^4$ & 50 \\
\tableline \tableline
\end{tabular}
\caption{Resolution of the numerical simulations. In our N-body simulations each model component (column one) is represented by a certain number of discrete particles (column two) of a given mass (column three). Column four provides the gravitational softening length that we adopt for each component.}
\label{tab:Res}
\end{center}
\end{table}

\begin{table}
\begin{center}
\begin{tabular}{l|l|l|l}
\tableline \tableline
label &  $v_{h}$ & $(x, y)_0$ & $(v_{x},v_{y},v_{z})_0$ \\ 
         &  (km/s)   & (kpc)            & (km/s)               \\ \tableline
vertical & 288 & (11.45, 0) & (-4.6, 0, -288) \\
prograde & 292 & (10.97, -1.96) & (12.4, 270, -110)  \\
retrograde & 291 & (11.05, 1.68) & (8.3, -270, -108) \\
\tableline \tableline
\end{tabular}
\caption{Properties of the substructure as it crosses the disk of the MW in the vertical, prograde, and retrograde simulations. Columns one and two show the simulation label and the substructure speed in the galactocentric restframe, respectively. Columns three and four provide the $x-y$ galactocentric coordinates and the velocity components of the density peak of the  substructure as it moves through the disk ($z=0$).  The centers of the MW disk, bulge, and halo are at rest at the coordinate origin.}
\label{tab:ICsMain}
\end{center}
\end{table}

In this work we study numerically the gravitational interaction between the disk of the MW and the orbiting substructure. In the main text we discuss three representative choices of orbital parameters: a vertical, a prograde, and a retrograde orbit, see Table~\ref{tab:ICsMain}. The inclination between the plane of the MW disk and the orbit plane of the substructure is approximately $90^\circ$, $20^\circ$, and $160^\circ$, respectively, in these cases. The prograde and retrograde orbits allow us to explore the impact of an orbiting substructure that co-rotates or counter-rotates with the majority of the stars in the MW disk. 

\begin{figure}
\begin{tabular}{c}
\includegraphics[width=80mm]{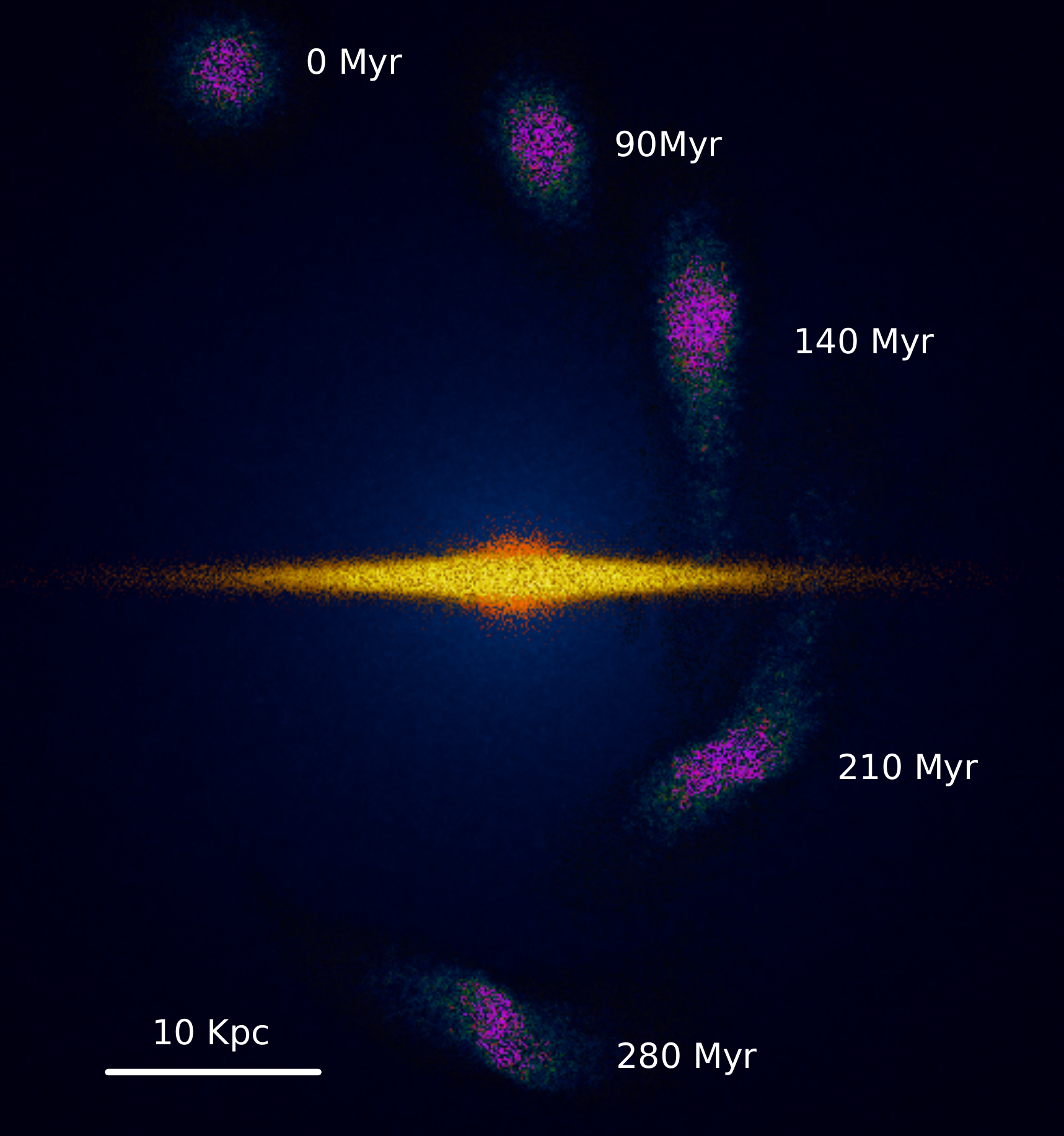}
\end{tabular}
\caption{A low mass substructure (purple) passing vertically through the stellar disk of the MW (yellow). Tidal forces deform the substructure noticeably, but do not destroy it. The simulated impact occurs 11 kpc from the Galactic Center at $t\sim{}190$ Myr.}
\label{fig:orbit}
\end{figure}

The initial position and velocity of the substructure put it on a collision course with the stellar disk of the MW. The impact occurs at 11 kpc from the Galactic Center with a speed of $\sim{}290$ km s$^{-1}$ , see Table~\ref{tab:ICsMain}. We describe the set-up of the MW -- substructure collisions in detail in Appendix \ref{app:ModelSetup}. We explore further orbital parameters in the appendix, finding little qualitative difference. We illustrate the vertical orbit of the substructure in Fig.~\ref{fig:orbit}.

We run our numerical simulations with PKDGRAV \citep{2001PhDT........21S}, the gravity solver of the TreeSPH code GASOLINE \citep{2004NewA....9..137W}. We adopt conservative values for the gravity opening angle (0.55) and the time stepping factor ($\eta=0.15$) in order to ensure an accurate integration of the equations of motions of the stellar and dark matter particles in our models.

We evolve the N-body realizations of the MW and the substructure in isolation for 500 Myrs to minimize non-equilibrium transients caused by the initial conditions. The structural and kinematical properties of our N-body model do not show significant evolution during this equilibration period indicating that the initial setup is indeed close to a self-consistent steady state. We subsequently integrate numerically the dynamical evolution of the combined MW -- substructure system. Our simulations span about 380 Myr of evolution, including about 190 Myr after the substructure passes through the MW disk.

\begin{figure}
\begin{tabular}{c}
\includegraphics[width=80mm]{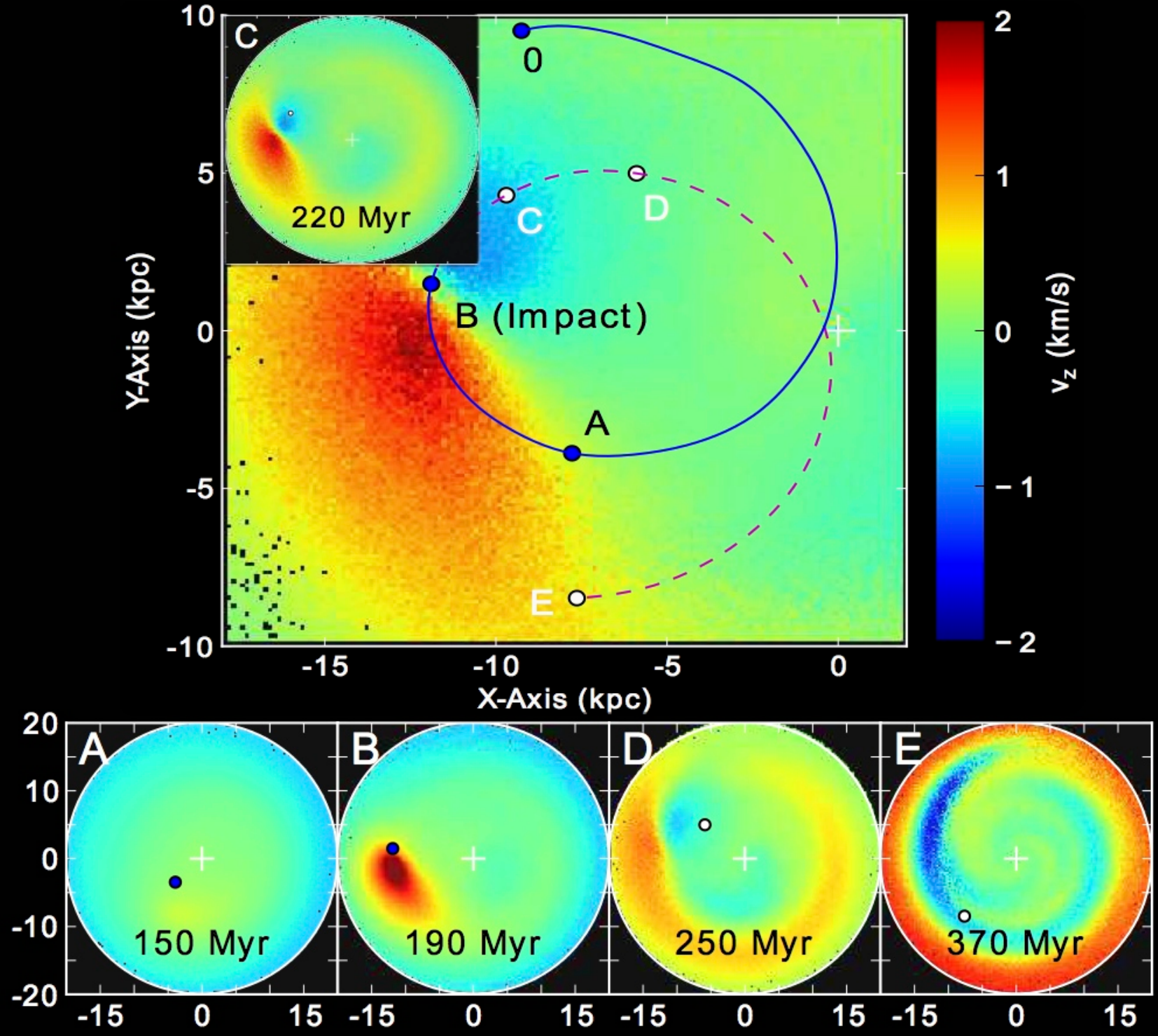}
\end{tabular}
\caption{Kinematic signature of a low mass substructure passing vertically through the disk of the MW. Each panel shows a velocity map of the face-on stellar disk of the MW model at a different time (see legend). The Galactic Center (white cross) is at $X=Y=0$. Panels A through E show the change in vertical velocity caused by the gravitational pull of the substructure in $500\times{}500$ pc$^2$ bins. Upward (downward) motions are shown in red (blue) colors. The blue (white) circle in each panel indicates the projected center of mass of the substructure when it is above (below) the MW disk plane. We show the position of the substructure in a frame co-rotating with the mean tangential velocity of stars at 8 kpc from the Galactic Center. The MW -- substructure interaction results in well-localized maxima and/or minima of the vertical velocity of disk stars, visible in panels A, B, C, and D.}
\label{fig:v_diff}
\end{figure}

\section{The kinematic signature of a low mass substructure passing through the MW}
\label{sect:CompSims}

We focus first upon the case of the substructure passing vertically through the disk. We show in Fig.~\ref{fig:v_diff} the effect of the substructure on the vertical motion $v_z$  of the stellar disk. Specifically, we show the change in $v_z$ after subtracting, particle by particle, the velocities from an otherwise identical reference simulation that does not include a substructure. As the substructure descends toward the disk, it gravitationally attracts part of the stellar disk below it, resulting in an upward motion (Fig.~\ref{fig:v_diff}A,B). Because the stellar disk rotates, the substructure exerts in general a downward force on a different part of the stellar disk after passing through the disk (Fig.~\ref{fig:v_diff}C,D). The result is that for a timescale of about 100 Myr the stellar disk either shows a well localized maximum of $v_z$, a minimum of $v_z$, or even both at the same time. The position of the velocity maximum (minimum) roughly tracks the projected position of the substructure when it is above (below) the disk. 

At later times (Fig.~\ref{fig:v_diff}E) the differential rotation winds up the localized velocity impulse resulting in an extended spiral-like pattern. In addition, the imparted velocity impulse may excite bending waves that start propagating across the disk and distorted the initial kinematic signal. Given that the solar neighborhood is likely stable against the buckling instability \citep{1994ApJ...425..551M, 2008gady.book.....B}, a conservative lower limit on the bending mode period is $\gtrsim{}100$ Myr. To obtain this lower limit we use the dispersion relation equation (7) in \cite{1971Ap&SS..14...52K} with a surface mass density of 50 $M_\odot$ pc$^{-2}$, a bending mode wave length of 2 kpc, and a vanishing in-plane velocity dispersion. The shearing of the disk is thus likely the dominant process by which the localized morphology of the kinematic imprint is erased.

The maximal velocity changes caused by the substructure are of the order of $\sim{}1-1.5$ km s$^{-1}$. This result agrees well with the prediction of equation (\ref{eq:DeltaV}) if we use the value $f\sim{}0.5$ appropriate for the given orbital parameters of the substructure (see appendix), identify $b$ with the scale radius of the substructure ($r_s=1.3$ kpc), and $M$ with the mass within the scale radius ($M_s=1.1\times{}10^{8}$ $M_\odot$). 

We can justify this choice of $b$ and $M$ as follows. Let the substructure have a dark matter density profile of NFW form $\rho(r)\propto{}r^{-1} \left(r_s+ r\right)^{-2}$ \citep{1996ApJ...462..563N}. The enclosed mass $M$ within a given radius $r$ increases linearly with $r$ for $r\sim{}r_s$, logarithmically with $r$ for $r\gg{}r_s$, and quadratically with $r$ for $r\ll{}r_s$. Hence, $M(<b)/b$, and thus the maximal change of $\Delta v_*$, is approximately constant for disk stars with impact parameter $b\sim{}r_s$, but decreases with increasing impact parameter for $b\gg{}r_s$ and decreases with decreasing impact parameter for $b\ll{}r_s$. Hence, the scale radius (here $r_s=1.3$ kpc) and the mass within the scale radius (here $M_s=1.1\times{}10^{8}$ $M_\odot$) of the passing substructure are the characteristic sizes and masses that upon inserting into equation (\ref{eq:DeltaV}) result in the largest velocity changes. We can account for a potential tidal truncation of the outer density profile by using $\min(r_s,r_t)$ and $\min(M_s,M_t)$ as characteristic sizes and masses, respectively. Here, $r_t$ is the tidal radius and $M_t$ is the mass within the tidal radius.

We simplify equation (\ref{eq:DeltaV}) further by making use of the NFW shape of the density profile. Simply calculus shows that $M_s/r_s = M_{\rm vir}/r_{\rm vir}\,g(c)$, where $g(c) = c[\ln(2)-0.5]/[\ln(1+c)-c/(1+c)]$ and $c=r_{\rm vir}/r_s$ is the concentration. $M_{\rm vir}$ is the virial mass of the substructure without tidal truncation or stripping, i.e., approximately the mass of the substructure when it first falls into the halo of the Galaxy. We note that tidal truncation of the NFW profile at $r>r_s$ has no bearing on the ratio $M_s/r_s$. Virial mass and virial radius are related via the chosen overdensity criterion (here 200 times the critical density at $z=0$). The term $g(c)$ lies between 1 and 2 for concentrations ranging from 1 to 25 ($g\sim{}1.7$ for the concentration $c=17$ of our simulated substructure). We can thus ignore any reasonable change of concentration with virial mass without significant loss of accuracy. Combining these various relations we can show that the maximal velocity change is of the order of
\begin{equation}
\max \Delta v_* \approx{}1-2\, \mathrm{km\,s^{-1}}  \left(\frac{M_{\rm vir}}{10^9\,M_\odot}\right)^{2/3} \left(\frac{400\,\mathrm{km\,s^{-1}}}{V}\right)f.
\label{eq:DeltaVNew}
\end{equation}
Hence, substructures with $M_{\rm vir}\sim{}10^9$ ($10^8$, $10^{10}$) $M_\odot$ result in typical velocity changes of the order of $1-2$ ($0.2-0.4$, $5-9$) km s$^{-1}$.

\begin{figure}
\begin{tabular}{c}
\includegraphics[width=80mm]{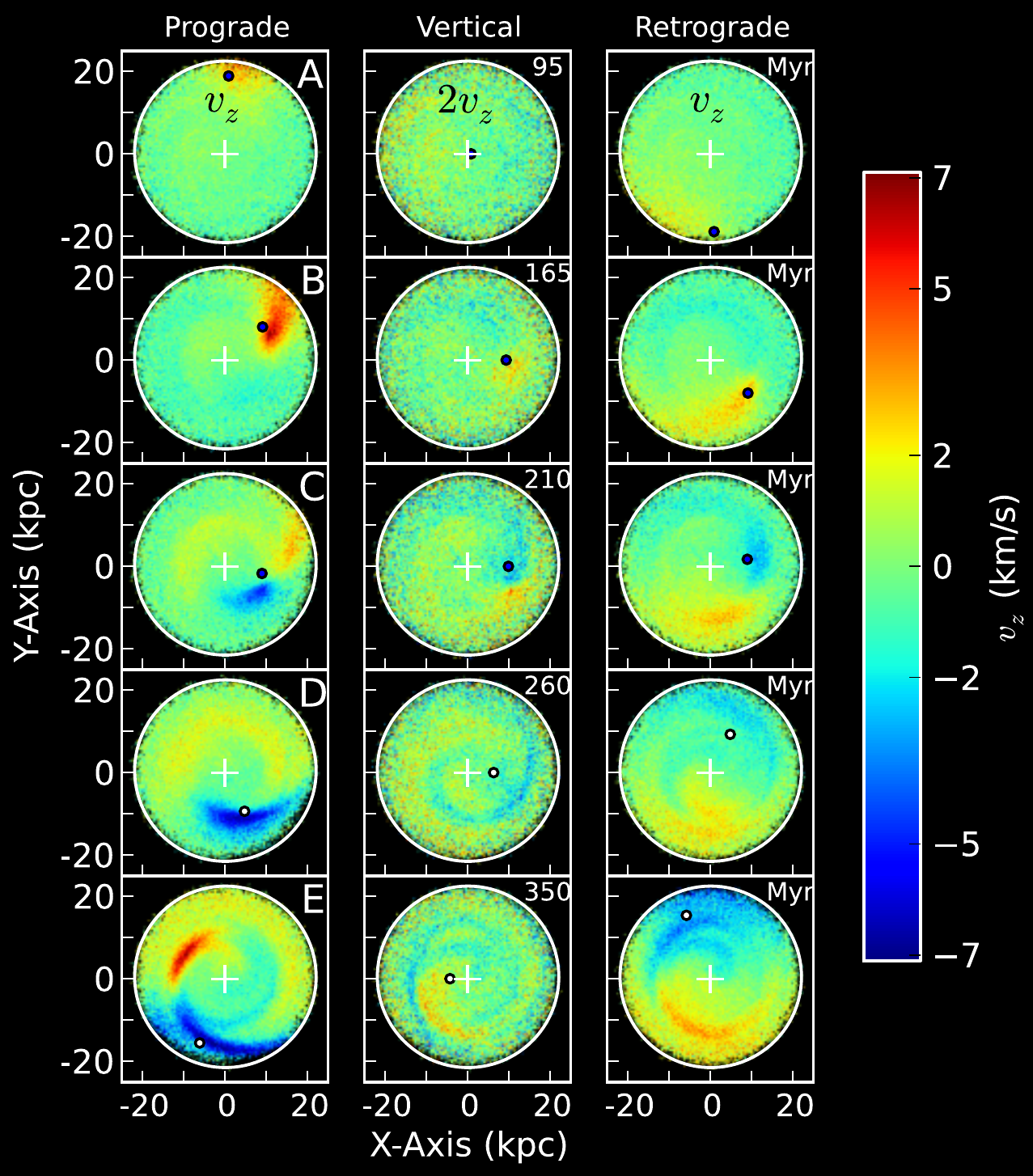}
\end{tabular}
\caption{Dependence of the kinematic signature on substructure orbit. The panels show the spatially averaged vertical velocity of MW disk stars for three different orbits and at five different times (see legend in the middle column). The spatial binning is $500\times{}500$ pc. The kinematic signature is strongest for a prograde orbit and weakest for a vertical orbit. In panels A-D (prograde) and B-C (retrograde and vertical case) the velocity disturbance is localized and traces approximately the projected position and the orbit of the substructure.}
\label{fig:v_z}
\end{figure}

The kinematic signature of a MW -- substructure interaction can be extracted \emph{without resorting to a reference simulation} by spatially binning the data. Fig.~\ref{fig:v_z} shows the average vertical velocity, $\langle{}v_z\rangle{}$, of disk stars in bins of $500\times{}500$ pc$^2$. In our simulations, bins at 8 kpc from the Galactic Center contain about 400 stellar particles. Consequently, the dispersion of $\langle{}v_z\rangle{}$ is lower than the dispersion of $v_z$ by a factor $\sqrt{400}=20$. As Fig.~\ref{fig:v_z} shows, the kinematic signature of the passing substructure is clearly visible in the binned vertical velocity. 

Fig.~\ref{fig:v_z} shows also the results for a prograde and a retrograde orbit of the substructure. Compared with the vertical orbit, both the prograde and the retrograde orbit enhance the strength of the kinematic signature of the MW -- substructure interaction. This is a consequence of the reduced vertical velocity of the substructure for an inclined orbit that results in a larger $f$ factor in equation (\ref{eq:DeltaVNew}), see also Appendix \ref{app:VelChange}. The increase of the kinematic signature is particularly dramatic for a prograde orbit as a result of the near matching of the orbital velocity of the satellite and the velocity of disk stars \citep{1972ApJ...178..623T}. The prograde passage of the substructure enhances the $v_z$ changes by more than a factor of 3 to about $\pm{}5$ km s$^{-1}$ at early and late times and to $\sim{}3$ km s$^{-1}$ during the collision of the substructure with the disk. In principle, a substructure on a co-rotating, grazing orbit could lead to even larger velocity changes.

The orbit of the substructure leaves tell-tale signatures in the spatial distribution of the vertical velocity changes, see Fig.~\ref{fig:v_z}C. For vertical orbits the kinematic signature is roughly circular in extent (at least until the shearing motion of the stellar disk distorts the shape). In contrast, a substructure on a prograde or retrograde orbit results in an aligned, elongated shape of the $v_z$ maximum. Hence, the measurement of high precision positions and velocities of stars across the MW disk may not merely enable the detection of dark matter substructures around the MW, but may also lead to a characterization of their orbital properties. We discuss additional runs with different orbital parameters in appendix \ref{app:AddSims}.

At late times (Fig.~\ref{fig:v_z}E) the velocity perturbances resist decay or may even grow in strength. Unfortunately, the shearing of the differentially rotating disk results in a large-scale spiral pattern which might be more difficult to categorize observationally than the localized kinematic imprint discussed above. Our worry is that the morphology, propagation, and strength of these late time disturbances are strongly influenced by the detailed structural properties of the whole MW system, i.e., the gradient of the rotation curve, the location of resonances, and the dynamics of the central bar.

\begin{figure}
\begin{tabular}{c}
\includegraphics[width=80mm]{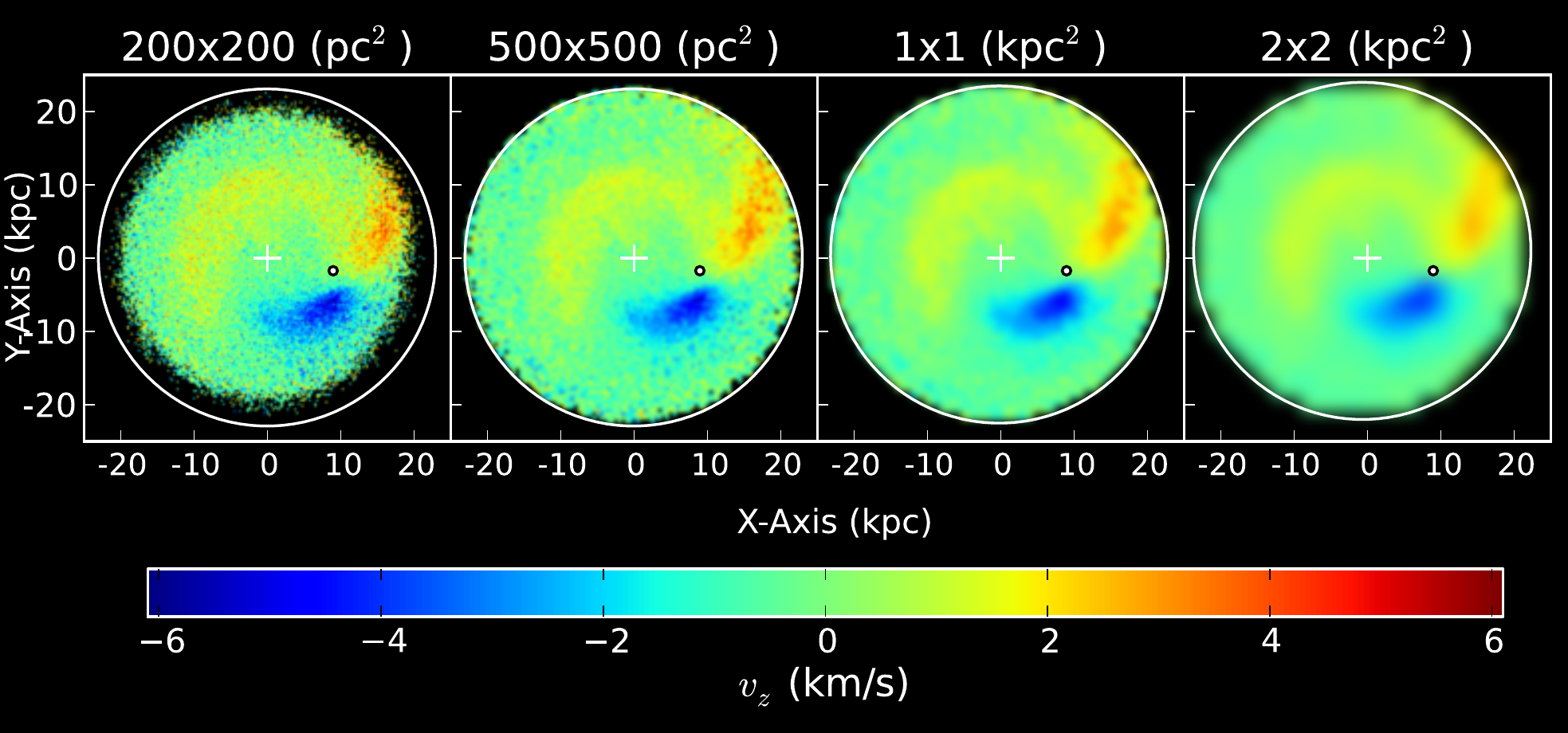}
\end{tabular}
\caption{Dependence of the kinematic signal on binning size. Each panel shows the leftmost panel C of Figure \ref{fig:v_z} (prograde orbit) for bins of different sizes (see legend). The original figure uses a binning of $500\times{}500$ pc$^{2}$, i.e., the same binning as shown in the left panel. The velocity minima and maxima are clearly visible independent of the chosen bin size.}
\label{fig:vz_500}
\end{figure}

In Fig.~\ref{fig:vz_500}, we test how the resolution of the spatial binning affects the detectability of the kinematic signal. The kinematic imprint of the substructure is clearly visible even for bins as large as 2 kpc$^2$.

\begin{figure}
\begin{tabular}{c}
\includegraphics[width=80mm]{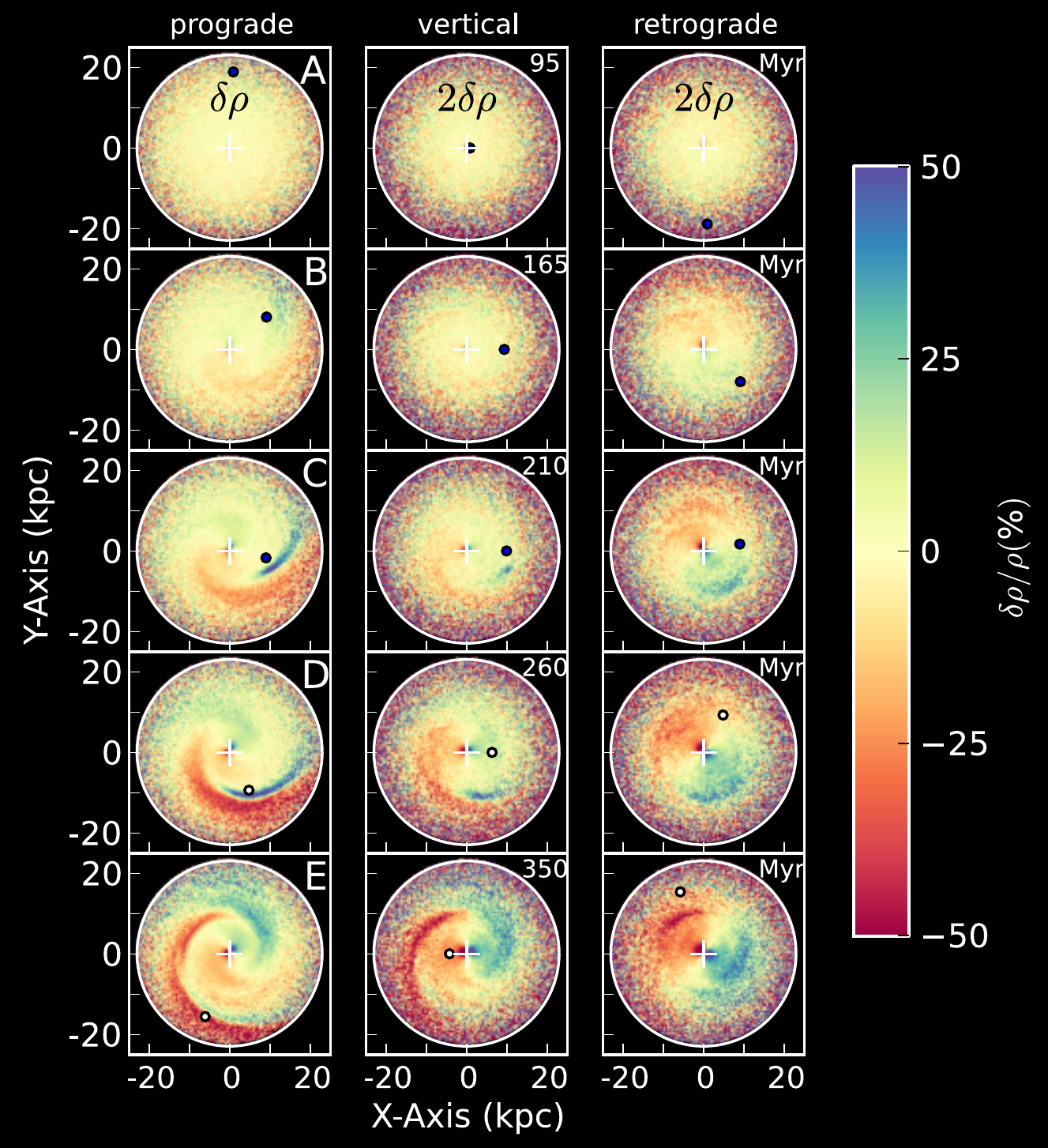}
\end{tabular}
\caption{Variations in the surface density of the MW disk that result from a low mass substructure that passes through the disk. The panels show maps of the fractional difference between the local surface density and the average surface density at a given galacto-centric radius. The middle and right panel show twice the density contrast. Low mass substructures passing through the MW disk induce disturbances of the stellar surface density that range from a few \% up to about 40\%. }
\label{fig:den}
\end{figure}

Aside from the kinematic imprint, the passage of the substructure also induces density variations that vary, depending on the orbit type, between 10\% and up to 40\%, see Fig.~\ref{fig:den}. The strongest density variations occur for a prograde passage of the substructure, the weakest for a vertical orbit. The disk develops, rather generically, a dipole in density with one side over dense and the other under dense. Substructures on prograde and vertical orbits also excite extended arm-like density enhancements \citep{1966ApJ...146..810J}, see Fig.~\ref{fig:den}C-E. Unfortunately, the generic morphology, the large spatial extent, and the non-uniqueness of the excitation mechanism will make it difficult to use the induced density variations as reliable tracers of substructures passing through the disk of the MW.

It is intriguing that these density perturbation induced by the passing substructure are only slightly weaker than those in spiral arms of observed nearby disk galaxies ($\sim{}15\%$ to $60\%$, e.g., \citealt{1995ApJ...447...82R}). Hence, in addition to massive satellite galaxies (e.g., \citealt{1972ApJ...178..623T}), low mass dark matter substructures (especially when on a prograde orbit) could be responsible for exciting coherent spiral structures on large scales in at least some disk galaxies.

\section{Rate and orbital parameters of substructures colliding with the MW disk}
\label{sect:Collisions}

A critical question that arises naturally is whether collisions with the MW disk are sufficiently frequent to allow for a realistic chance of detecting sub-halos in future surveys. To answer this question we compute in this section the number of substructures that cross the disk over the life-time of the kinematic imprint ($\sim{}100-200$ Myr).

The substructure -- disk collision rate scales with the mean speed $\langle{}v_{\rm h}\rangle{}$ of substructures, the number density $\langle{}n_{\rm h}\rangle$ of sub-substructures, and the geometric cross section $\sigma_g=\pi R^2$ of the disk with radius $R$
\begin{equation}
\Gamma_{\rm c}\approx{}2\,\langle{}v_{\rm h}\rangle{}\,\langle{}n_{h}\rangle\,\sigma_g.
\label{eq:Gammac}
\end{equation}
The prefactor accounts for the likely scenario that the substructure crosses the disk twice per pericentric passage.

We calculate the average number density of substructures directly from the Aquarius simulation suite \citep{2008MNRAS.391.1685S}. Aquarius is a set of high-resolution N-body simulations of a MW like dark matter halo. The mass function of substructures is provided in equation (4) of \cite{2008MNRAS.391.1685S}. The number density of substructures with masses above $M_{\rm min}$ within a $r_{50}=430$ kpc radius around the main halo is
\[
N(>M_{\rm min}) = 264\left(\frac{10^8\,M_\odot}{M_{\rm min}}\right)^{0.9}.
\]
$M_{\rm min}$ refers to the gravitationally bound mass of substructures in the Aquarius simulation at $z=0$. As a consequence of tidal stripping this mass is significantly smaller (a factor 10 is typical\footnote{The Hill radius of a $10^8$ $M_\odot$ point mass at $R=20$ kpc from the center of the Galaxy with $M_{\rm tot}(<20\,\textrm{kpc})\sim{}2\times{}10^{11}$ $M_\odot$ is about 1 kpc. A $M_{\rm vir}=10^9$ $M_\odot$ dark matter halo with an NFW profile has a scale radius that roughly coincides with this Hill radius. Hence, it will likely be tidally stripped down to the mass within the Hill radius, i.e., down to $\sim10^8$ $M_\odot$. This approximate correspondence between scale radius and tidal radius holds independent of the mass of the substructure (at fixed $c$, $R$, and $M_{\rm tot}(<R)$),because both radii scale with their respective masses in the same way $r\propto{}M^{1/3}$.}) than the virial mass of the substructure when it first entered the main halo.

\begin{figure*}
\begin{tabular}{ccc}
\includegraphics[width=53.33mm]{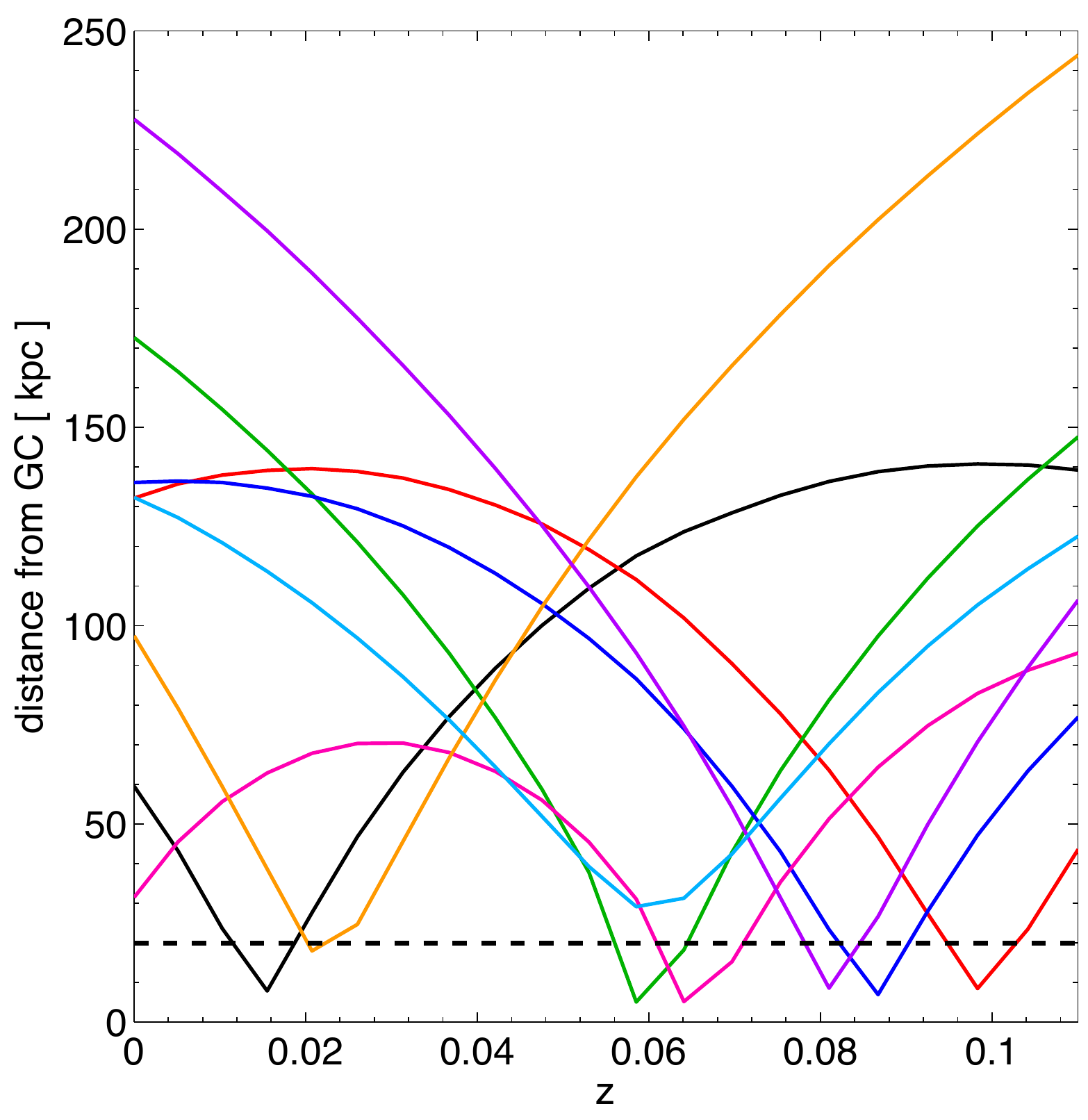} & 
\includegraphics[width=55mm]{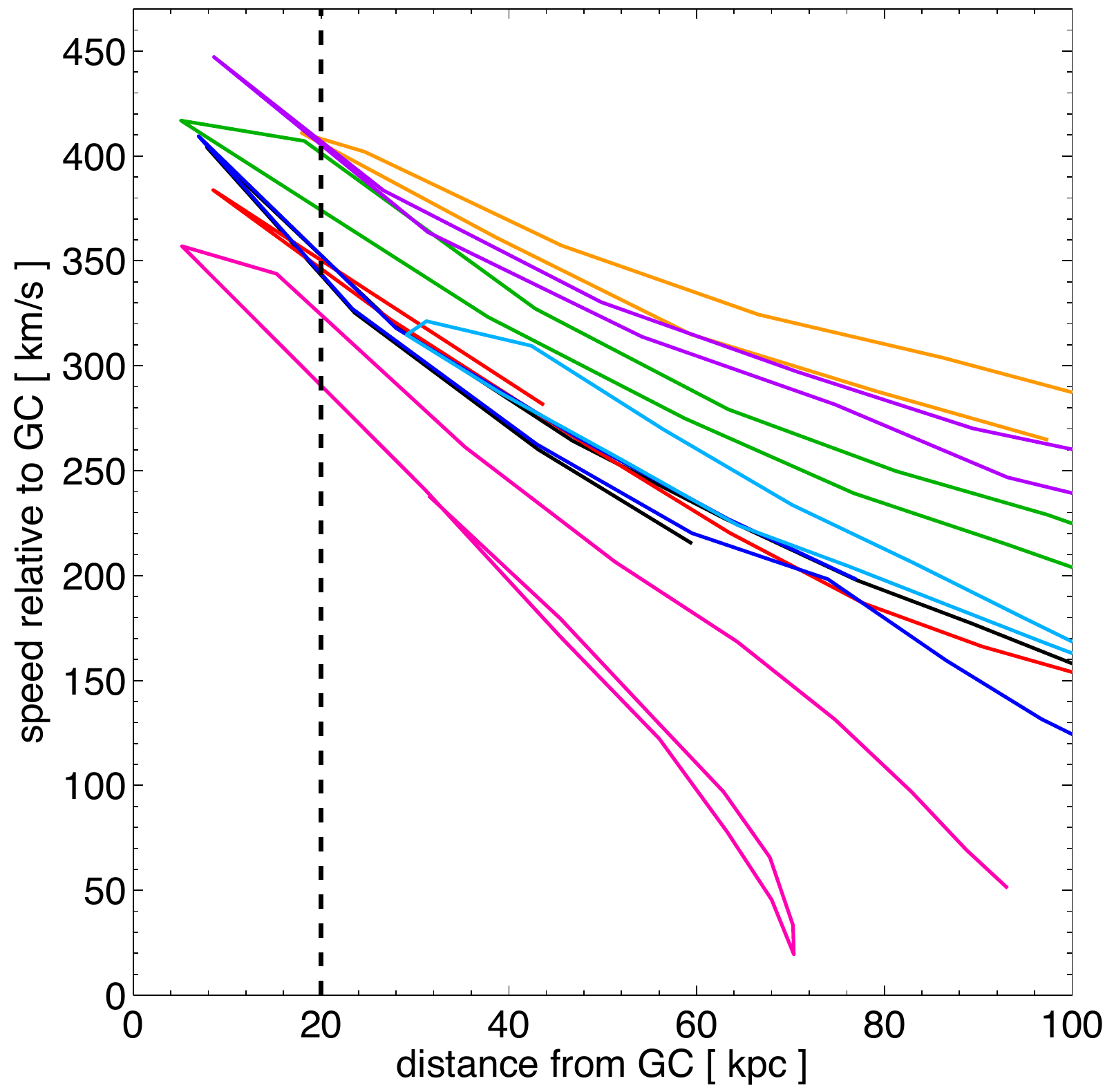} &
\includegraphics[width=55mm]{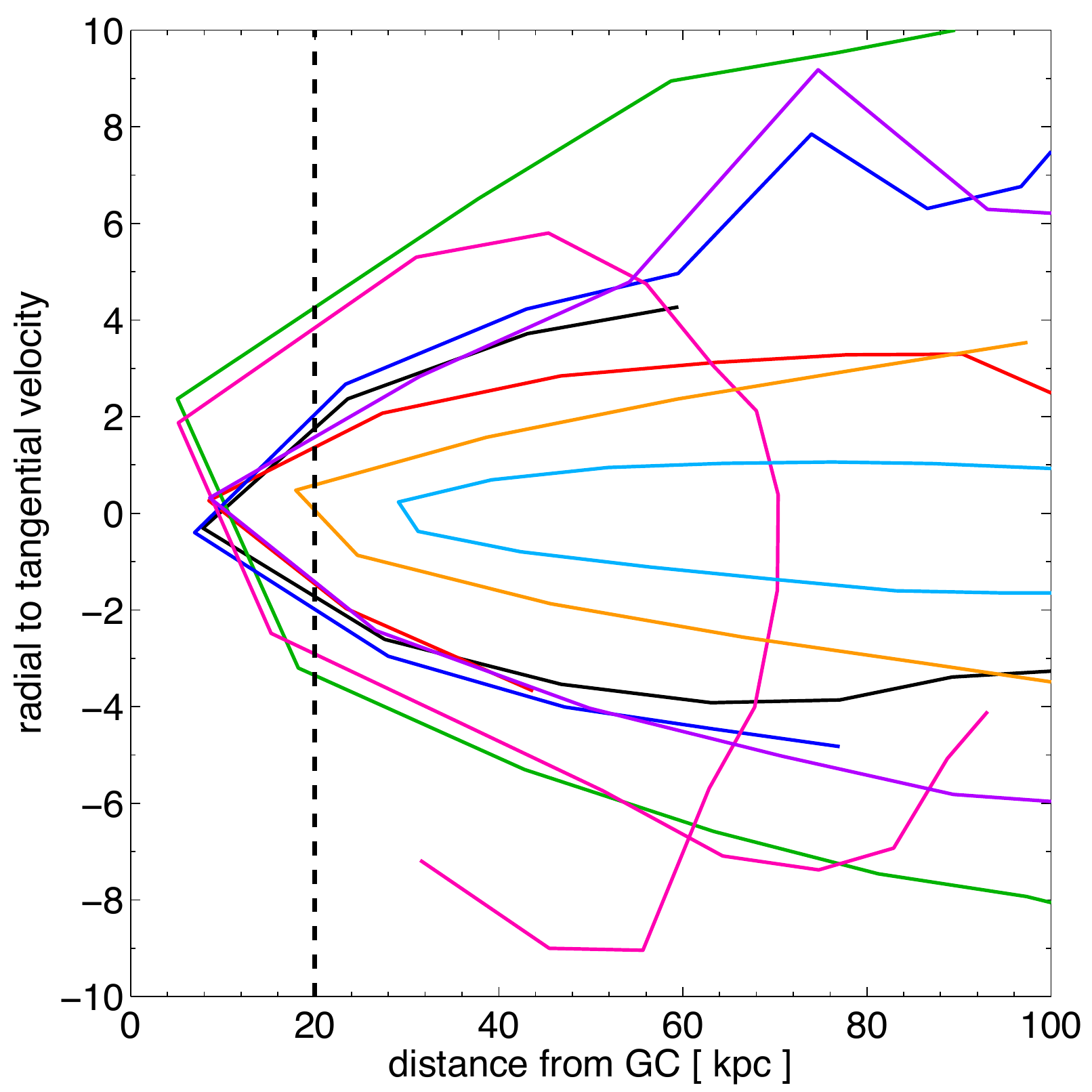}
\end{tabular}
\caption{Orbits of selected substructures in the Via-Lactea I simulation. Substructures are selected based on their mass ($>10^8$ $M_\odot$), their peak maximum rotation velocity ($>20$ km/s), and their distance from the Galactic Center ($<30$ kpc). Seven substructures match these criteria and enter the innermost 20 kpc of the main halo (dashed lines). (Left panel) Distance from the Galactic Center vs redshift. About 1 selected substructure crosses the disk of the MW per 100 Myr. (Middle panel) Speed of the selected substructures vs distance from the Galactic Center. Substructures intersecting the MW disk have typical speeds of $\sim{}300-450$ km/s. (Right panel) Ratio of radial to tangential velocity vs distance from Galactic Center. The ratio is negative if the substructure moves inward and positive if it moves outward. Substructures intersecting the MW disk have typical radial to tangential velocity ratios in the range of $-3$ to $+3$.}
\label{fig:CollOrbits}
\end{figure*}

\cite{2008MNRAS.391.1685S} show that the number density of substructures can be well fit with an Einasto profile \citep{1965TrAlm...5...87E} and that the shape parameters are independent of the substructure mass. Using the fit parameters provided in section 3.2 of  \cite{2008MNRAS.391.1685S} we find that the mean interior number density of substructures of a given mass increases by a factor $\sim{}20$ between $r_{50}=430$ kpc and $r=20$ kpc. Hence, the mean density of substructures with masses $>M_{\rm min}$ and within $r=20$ kpc is approximately
\[
\langle{}n_{\rm h}\rangle \approx{} 1.7\times{}10^{-5} \left(\frac{10^8\,M_\odot}{M_{\rm min}}\right)^{0.9} \text{ kpc}^{-3}.
\]
The mean density is not a very sensitive function of the enclosing radius, e.g., choosing $r=5$ kpc would increase it by only 40\%.

The scale length of the MW disk is 2-3 kpc and its full spatial extent is about 15-20 kpc~\citep{2008gady.book.....B}. Substructures that enter the central 20 kpc of the MW halo have typical speeds of $\sim{}400$ km s$^{-1}$ as we will demonstrate later in this section. Inserting $\langle{}v_{\rm h}\rangle{}=400$ km s$^{-1}$, $R=20$ kpc, and $\langle{}n_{\rm h}\rangle$ into equation (\ref{eq:Gammac}) we find
\begin{equation}
\Gamma_{\rm c}\approx{}1.8 \left(\frac{10^8\,M_\odot}{M_{\rm min}}\right)^{0.9} \text{ per 100 Myr}.
\label{eq:GammacComputed}
\end{equation}
Hence, we expect several disk crossings per dynamical time of the MW disk for substructures with tidal masses above $10^8$ $M_\odot$ and virial masses above $10^9$ $M_\odot$.

As a consistency check we also estimate the collision rate from publicly available substructure orbits provided by the Via-Lactea project\footnote{see \url{http://www.ucolick.org/~diemand/vl/}}. Via-Lactea I is a cosmological N-body simulation that follows the formation of a MW like dark matter halo and resolves over 6000 substructures with a peak circular velocity above 5 km/s. We identify 8 substructures from Via-Lactea I that satisfy all three of the following conditions at some redshift $z\leq{}0.1$ (i.e., within the past 1.3 Gyr): (i) a bound mass above $10^8$ $M_\odot$, (ii) a peak maximum rotation velocity above 20 km/s (to ensure that the virial mass at infall was $\gtrsim{}10^9$ $M_\odot$), and (iii) a position within 30 kpc from the Galactic Center, see left panel of Fig.~\ref{fig:CollOrbits}. Seven out of the 8 substructures have a pericentric distance to the Galactic Center of less than 20 kpc and, hence, would intersect the disk of the MW. Given that most of these substructures would cross the disk twice we arrive at $\Gamma_{\rm c}\sim{}1.1\pm{}0.4$ per 100 Myr, in reasonable agreement with our previous estimate (\ref{eq:GammacComputed}).

We note that these estimates are only approximate. The mass of the MW halo and, thus, the expected abundance of dark matter substructures are constrained observationally only to within a factor of a few (e.g., \citealt{2013ApJ...768..140B}). In addition, our estimates are based on pure dark matter simulations that neglect baryonic processes. Baryons can enhance the collision rate via adiabatic contraction and gravitational focusing. Conversely, the destruction of substructures in previous dynamical interactions with the disk could reduce the interaction rate by a factor 2-3 \citep{2010ApJ...709.1138D}.

The orbits provided by the Via-Lactea simulation allow us to constrain the typical speeds and orbital parameters of substructures passing through the disk of the MW. In the middle panel of Fig.~\ref{fig:CollOrbits} we show the speed as function of distance from the Galactic Center for the 8 selected substructures (see above). Typical speeds are $300-450$ km/s depending on the chosen substructure and on the orientation between the orbit and the MW disk. The right panel of Fig.~\ref{fig:CollOrbits} shows that the radial to tangential velocity of substructures has a broad distribution ranging from nearly tangential motions (ratio $\sim{}0$), to strongly radial motions (absolute value of the ratio $\sim{}3$). We note that the substructure orbits chosen in this work reflect the range of typical speeds and radial to tangential velocities found in the Via-Lactea cosmological simulation, see Table~\ref{tab:ICsMain} and Appendix \ref{app:ModelSetup}.

\section{Detecting low mass substructures with Gaia}
\label{sect:Gaia}

The upcoming astrometric mission Gaia will provide positions and motions for over a billion MW stars, observe objects out to 1 Mpc, and at a micro-arcsecond ($\mu as$) precision \citep{2001A&A...369..339P}. Gaia is in fact ideally suited to search for the kinematic signatures of starless substructures orbiting the MW as we now demonstrate.

\subsection{Measuring the mean velocity of disk stars with Gaia}

Gaia will be able to measure parallaxes to a precision\footnote{see Gaia Science Performance at \url{http://www.cosmos.esa.int/web/gaia/science-performance}}  of $26$ $\mu as$ and proper motions to $14$ $\mu as$ yr$^{-1}$ for stars with an apparent magnitude brighter than 15 in the $G$-band (which is the main photometric band for Gaia).  We estimate that Gaia should observe about 10$^{5}$ disk stars per kpc$^2$ with a parallax error less than $20$ $\mu as$ (and a corresponding distance error of $<$ $10$\%) at a distance of 5 kpc. This surface density is sufficient to detect passing substructures of mass $\gtrsim{}10^{8}$ $M_\odot$.

To arrive at this estimate we use the code Galaxia~\citep{2011ascl.soft01007S} to create a realistic mock catalog of MW stars. The code returns the absolute magnitude of each star in the $V$ and the $I$ band, the distance to the star, and the extinction due to dust. We convert absolute magnitudes into apparent magnitudes using the known distances and dust extinctions and then use the fitting formulae provided by \cite{2010A&A...523A..48J} to estimate the parallax error of each star. 

\begin{figure}
\includegraphics[width=80mm]{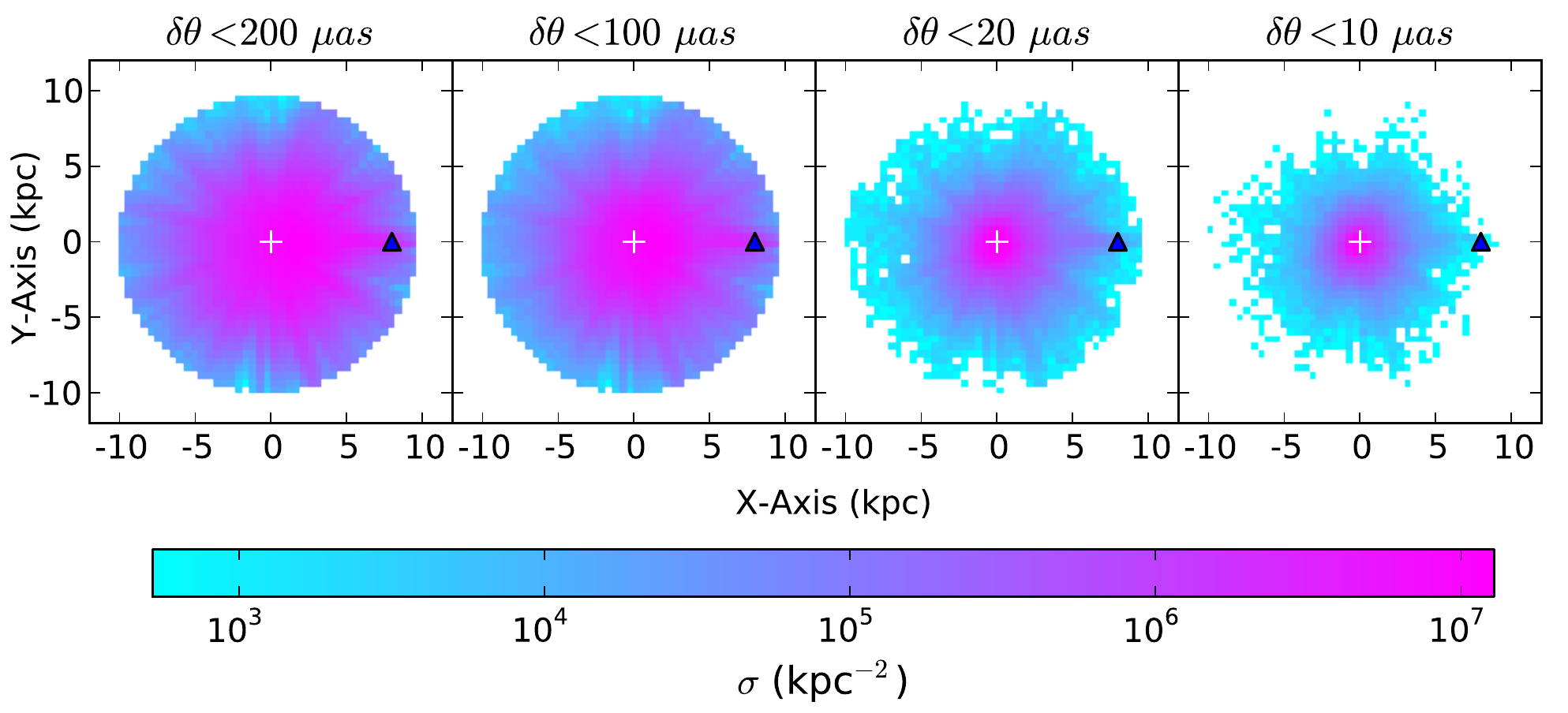}
\caption{Surface density of stars for which Gaia can measure a parallax to better than 
200, 100, 20 and 10 $\mu as$. The limits of the parallax error correspond to apparent $G$ band magnitude limits of 19.7, 18.5, 15.1, and 13.6 (from left to right).
 The white cross marks the location of the sun. The blue triangle represents the Galactic Center.  }
\label{fig:sigma_p}
\end{figure}

In Fig.~\ref{fig:sigma_p}, we show the surface density of stars for which Gaia is able to measure parallaxes to better than a specified parallax error. We expect that Gaia will observe more than 10$^{5}$ (10$^{7}$, 10$^{3}$) disk stars per kpc$^{2}$ at a 5 (2.5, 10) kpc distance with a parallax error better than 20 $\mu$as.

\boldtext{The parallax error determines both the distance error and the velocity error transverse to the line of sight for a particular star and thus has a strong impact on the ability of Gaia to detect low mass substructures. Given $d[\mathrm{pc}]=1/\theta[\mathrm{''}]$ and $v[\mathrm{km/s}] = 4.74\,d[\mathrm{pc}]\,\mu[\mathrm{''/yr}]$, the relative distance error $\delta d / d$ for a star equals the relative parallax error $\delta{}\theta / \theta$ and the velocity error $\delta v[\mathrm{km/s}]$ scales as $\lesssim{}4.74\,d[\mathrm{pc}]\,\delta{}\mu[\mathrm{''/yr}]$ + $\delta{}v[\mathrm{km/s}]\,\delta{}d/d$. Given a typical transverse velocity of $\sim{}50-80$ km s$^{-1}$ of a star at $d=5$ kpc from the Sun\footnote{\boldtext{To obtain this estimate we model the disk of the Milky Way as a cold stellar disk rotating at $200$ km s$^{-1}$ with the Sun at $8$ kpc from the Galactic Center. The relative transverse velocity of stars at 5 kpc from the Sun in a Galactic rest frame varies between 0 km s$^{-1}$ and $129$ km s$^{-1}$, with an angle average velocity of $\sim{}66$ km s$^{-1}$.}}, a parallax error $\delta{}\theta=20\mu as$, and a corresponding proper motion error$^3$ $\delta{}\mu[\mathrm{mas/yr}] = 0.526\,\delta{}\theta{}[\mathrm{mas}]$, we find that $\delta v\sim{}5-8$ km s$^{-1}$. The precision on distance and transverse velocity degrade quickly with distance\footnote{\boldtext{Beyond parallax, we have not considered better techniques which could improve measuring the location of a star along the line of sight. For example, given that the angular velocity of a star changes strongly as a function of radius from the Galactic Center one could use the angular velocity to help locate the position of a star along the line of sight which may significantly reduce distance errors. We will leave such possibilities for the future.}}, however. For instance, they are 125 pc and $\sim{}1-2$ km s$^{-1}$ for stars at 2.5 kpc distance, but 2 kpc and $\sim{}20-40$ km s$^{-1}$ for stars 10 kpc from the Sun.}
  
In the simplest scenario (Poisson noise) the error of the spatially averaged velocity $\bar{v}$ scales with the observed stellar velocity dispersion $\sigma_s$ and with one over the square root of the number of stars N in the given spatial bin, i.e.,
\beq
\delta\bar{v}=\frac{\sigma_s}{\sqrt{N}}
\eeq
\boldtext{Both the intrinsic stellar velocity dispersion of stars in the Milky Way disk and the transverse velocity errors contribute to $\sigma_s$. For stars at $d<10$ kpc $\sigma_s$ is dominated by the intrinsic velocity dispersion  ($\sim{}20-30$ km s$^{-1}$ in the solar neighborhood), while for $d\sim{}10$ kpc the velocity errors and the intrinsic velocity dispersion contribute about equally. Hence, unless additional complications arise (see below) the average velocity $\bar{v}$ can be measured to better than $\sim{}1$ km s$^{-1}$ accuracy if the spatial bins contain more than 10$^{3}$ stars. This accuracy is sufficient to detect any 10$^9$ $M_\odot$ substructure that passes through the MW disk within a 10 kpc distance.} In addition, 10$^8$ $M_\odot$ substructures can be detected if they collide with the MW disk on prograde orbits. Furthermore, with 10$^{5}$ stars per bin $\bar{v}$ can be constrained to better than 0.08 km s$^{-1}$, which is sufficient to detect any 10$^8$ $M_\odot$ substructure (as well as 10$^7$ $M_\odot$ substructures on prograde orbits) within a 5 kpc distance.

\subsection{Caveats}

A variety of complications could potentially diminish the sensitivity of the proposed detection method. First, nearby stars may have correlated velocities, reducing the effective number of independent velocity measurements. However, outside of stellar clusters and associations, this effect should be small and, given the size of the Gaia data set, should not constitute a limiting factor for measuring accurate spatially averaged velocities. 

A second and potentially more serious issue is whether Gaia can distinguish $\sim{}$km s$^{-1}$ velocity disturbances caused by a passing substructure from fluctuations caused by other sources. Focusing on the vertical velocity simplifies matters because spiral arms typically excite velocity variations in the plane of the disk \citep{2008gady.book.....B}. Furthermore, stellar density waves in the vertical direction should die out on reasonably fast time scales \citep{2012ApJ...750L..41W}. Most importantly, however, the unique morphology of the highly localized $v_z$ maxima and/or minima clearly distinguishes the changes that result from a passing substructure from the changes caused by a density wave. 

Third, molecular clouds, star clusters, globular clusters, and satellite galaxies may affect the velocities of disk stars. However, these objects are visible, while the telltale sign of low mass dark matter substructure is a perturbation of the disk without a visible counterpart. Globular clusters are also not a concern given their low masses and their spatial distribution that is concentrated toward the Galactic Center.

Our simulations focused on a single encounter between a dark substructure and the disk of the MW. A potential concern is that the long-levity of bending and spiral modes excited by previous passages may mask the (weak) kinematic imprint of subsequent collisions. We hope to address this important question in future work.

\section{Summary and conclusions}
\label{sect:Summary}

We have studied the impact of low mass dark matter substructures as they pass through the disk of the MW with the help of high resolution numerical simulations. Our main findings are as follows

\begin{itemize}
\item The passage of a substructure results in distinct, coherent variations in the vertical velocities of disk stars. The morphology of the kinematic signal clearly distinguishes it from other disturbances such as spiral waves. The spatial size of the signature is of the order of the scale radius of the passing substructure. The strength of the kinematic disturbance scales with the mass contained within the scale radius of the substructure.

\item For a low mass substructure ($M_{\rm vir}\sim{}10^9$ $M_\odot$, $M(<r_s)\sim{}10^8$ $M_\odot$) the velocity changes are of the order of one to several km s$^{-1}$, depending on the orbit of the substructure. A prograde orbit results in the strongest signal, a vertical orbit in the weakest signal. The kinematic signature is coherent on scales of a few kpc.

\item If CDM theory is correct, we expect about 2-20 dark matter substructures with virial masses $\gtrsim{}10^8-10^{9}$ $M_\odot$ to collide with the disk of the MW per dynamical time ($\sim{}200$ Myr). Given the long lifetimes ($\sim{}100$ Myr) of the kinematic signature of a substructure passage through the MW disk, we expect potentially several such signatures be present at any given time.

\item The Gaia space mission is ideally suited to search for these kinematic signatures given its unprecedented accuracy in distance and velocity measurements, its large spatial coverage and sample size. Data from the Gaia mission should allow to detect  starless dark matter substructure with masses $\sim{}10^8-10^9$ $M_\odot$. Whether substructures of even lower masses can be detected in the same way depends on the presently unknown strength and properties of low-level vertical velocity perturbations across the MW disk. 

\end{itemize}

Interestingly, recent observations indicate significant variations of the mean vertical velocity at moderate heights above and below the disk plane \citep{2012ApJ...750L..41W,2013MNRAS.436..101W, 2013ApJ...777L...5C}. The origin of this kinematic feature is unknown, but it may well be a density wave excited by an external perturber \citep{2012ApJ...750L..41W, 2013ApJ...777L...5C, 2014MNRAS.440.1971W}. The observed variations are of the order of $\sim{}10$ km s$^{-1}$ at a $\gtrsim{}$ kpc height above/below the stellar disk. They are significantly weaker ($\lesssim{}1-2\sim{}$ km s$^{-1}$), however, at lower altitudes and thus potentially reflect the kinematic imprint of a low mass dark matter substructure passing through the MW disk. A crucial next step in understanding the origin of the kinematic feature will be to map the large scale ($>$kpc) morphology of the feature and to compare it with theoretical predictions, such as those provided in Fig.~\ref{fig:v_z}.

The detection of individual low-mass substructures orbiting the MW will complement the estimates of cumulative substructure fractions in distant galaxies based on gravitational lensing measurements \citep{1998MNRAS.295..587M,2002ApJ...572...25D}. In addition, the high precision astrometric data from Gaia will hopefully allow to put constraints on the orbital properties and the mass function of the starless substructures. As such, the proposed experiment will provide the basis for a crucial test of the CDM paradigm, leading potentially to new insights into the nature of dark matter and the physics of galaxy formation in low mass halos.

\section*{Acknowledgments}
We thank the anonymous referee for their valuable comments that helped improving the quality and clarity of the paper. We are grateful to Joe Silk, Matt Lehnert, and Matt McQuinn for constructive comments. We also thank Larry Widrow for providing us with the latest version of the GalactICS code and with the parameters of the MW model. R.F. acknowledges support for this work by NASA through Hubble Fellowship grant HF-51304.01-A awarded by the Space Telescope Science Institute, which is operated by the Association of Universities for Research in Astronomy, Inc., for NASA, under contract NAS 5-26555. The research of D.S. has been supported at IAP by  the ERC project  267117 (DARK) hosted by Universit\'e Pierre et Marie Curie - Paris 6. This work made extensive use of the NASA Astrophysics Data System and {\tt arXiv.org} preprint server.


\appendix

\section{Velocity change induced by a passing substructure}
\label{app:VelChange}

The gravitational pull of a substructure orbiting in the MW halo affects the velocity of stars in the stellar disk. We can estimate the magnitude and spatial extent of this perturbation using the free-particle approximation of the disk star motion. Our specific setup is as follows.

We choose a coordinate system in which $z=0$ is the mid-plane of the disk and the star is at rest at position $\vec{r}_*$. The orbiting substructure moves at constant velocity and passes through the coordinate origin at $t=0$, i.e., $\vec{r}_s(t) = \vec{v}_st$. The impact parameter $\vec{b}=\min_t\left(\vec{r_*}-\vec{r}_s\right)$ of the interaction is 
\[
\vec{b} = \vec{r}_* - \vec{v}_s\frac{\vec{r}_*\mathbf{\cdot}\vec{v}_s}{v_s^2}, \textrm{with } b^2 = r_*^2 - \frac{(\vec{r}_*\mathbf{\cdot}\vec{v}_s)^2}{v_s^2}
\]

The unbound gravitational interaction between two point masses is covered in standard textbooks (e.g., \citealt{2008gady.book.....B}). The gravitational encounter between a disk star with mass $M_*$ and a point-like substructure with mass $M_s$ results in the change $\Delta\vec{V}$ of their relative velocity $\vec{V} = \vec{v}_* - \vec{v}_s$ with
\[
\vert\Delta{}\vec{V}_\perp\vert = 2\,V\frac{b/b_{90}}{1 + b^2/b_{90}^2}, \textrm{ and }\vert\Delta{}\vec{V}_\parallel\vert=2\,V\frac{1}{1 + b^2/b_{90}^2}.
\]
Here, $b_{90} = G(M_h+M_*)/V^2$ is the impact parameter that leads to a $90^{\circ}$ deflection and $V=\vert\vec{V}\vert=v_s$. Note that $\vec{V}_\perp$ and $\Delta{}\vec{V}_\parallel$ point in the direction opposite to $\vec{b}$ and $\vec{V}$, respectively. The velocity change of the disk star is $\Delta{}\vec{v}_* = \frac{M_h}{M_h+M_*}\Delta{}\vec{V} \approx{} \Delta{}\vec{V}$ (since $M_*\ll{}M_s$), i.e., 
\begin{flalign}
\Delta{}\vec{v}_{*} &\approx{} \frac{2\,v_s}{1+b^2/b_{90}^2}\left(\frac{\vec{v}_s}{v_s} - \frac{\vec{b}}{b_{90}}\right) \notag \\
                              &\approx{} -\frac{2\,G\,M_s}{v_s} \frac{\vec{b}}{b^2}
\label{eq:dvs}
\end{flalign}
The latter approximation is valid in the limit $b\gg{}b_{90}\approx{} 0.048\, \mathrm{kpc} \left(\frac{M_s}{10^9\,M_\odot}\right) \left( \frac{v_s}{300\,\mathrm{km}\,\mathrm{s}^{-1}}\right)^{-2}$.

\begin{figure*}
\begin{tabular}{cc}
\includegraphics[width=80mm]{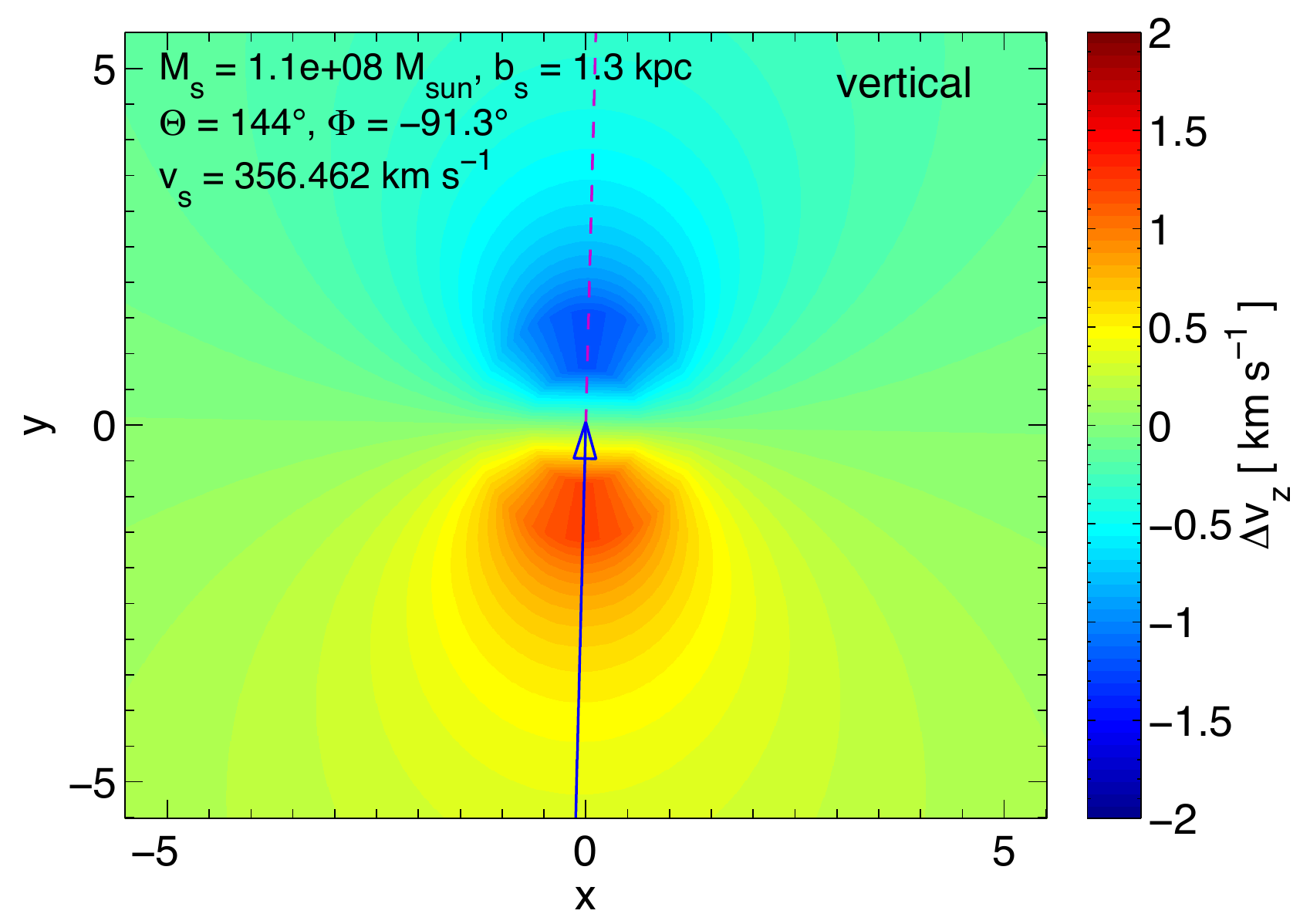} &
\includegraphics[width=80mm]{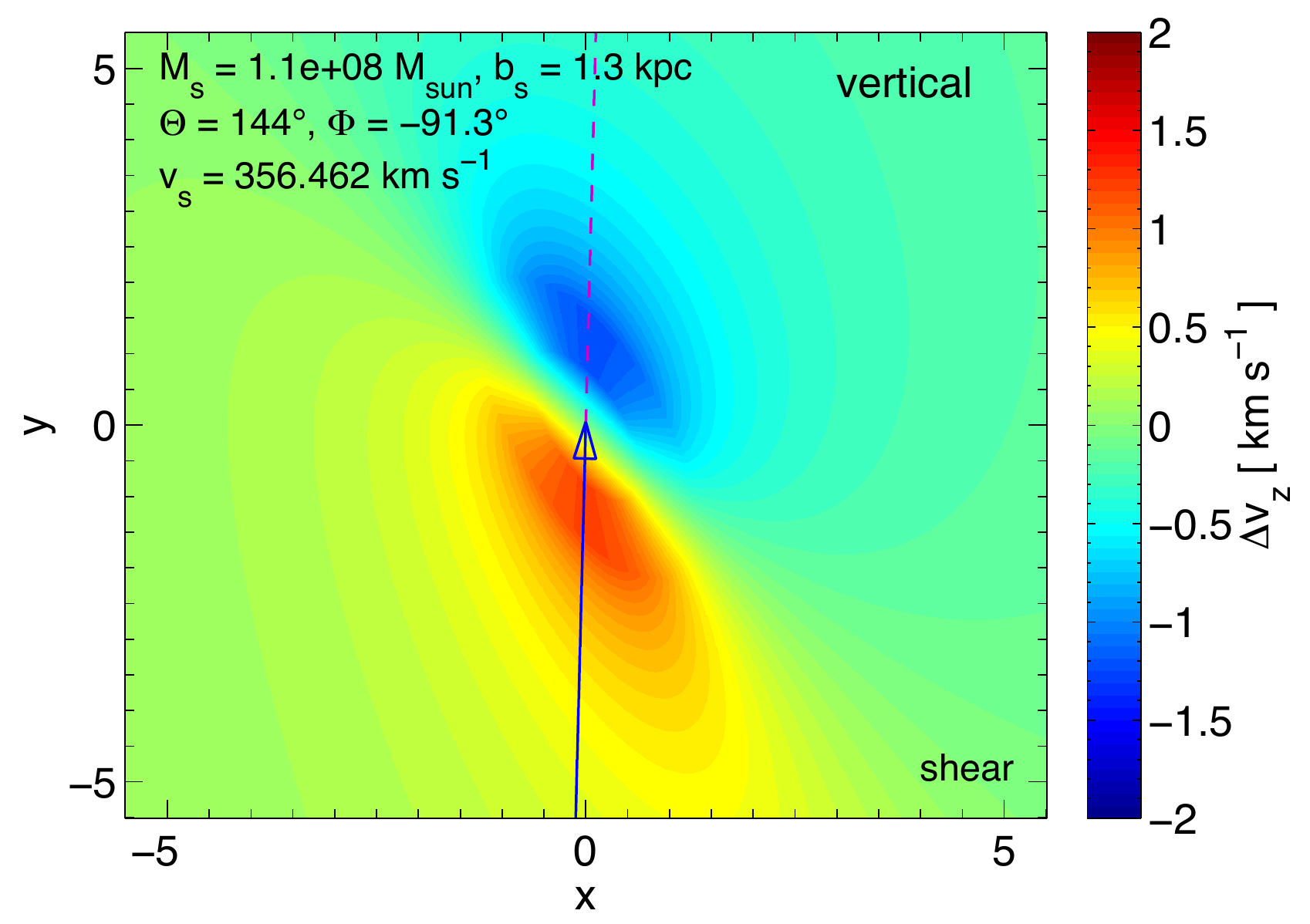}
\end{tabular}
\caption{Perturbations of the vertical velocity of disk stars caused by an orbiting substructure as predicted by equation (\ref{eq:dvs}). The $x$-$y$ plane corresponds to the stellar disk in its local standard-of-rest. The substructure passes through the disk at $x=y=0$. 
(Left panel) Without shearing of the stellar disk. (Right panel) Shearing of $v_y$ with a gradient of 21 km s$^{-1}$ kpc$^{-1}$ along the x-axis. The panel shows the distorted velocity map approx. 50 Myr after the substructure crosses the disk. The total mass of the substructure is $M_s=1.1\times{}10^8$ $M_\odot$. We rescale this mass for small impact parameters  $b\leq{}b_s=1.3$ kpc to account for the extended size of the substructure. Specifically, we adopt a continuous scaling $M_s\propto{}b$ for $0.5\,b_s\leq{}b\leq{}b_s$ and $M_s\propto{}b^2$ for $b\leq{}0.5\,b_s$. The velocity angles $\Theta$ and $\Phi$ and the relative speed $v_s$ (see legend) mimic the vertical orbit of the substructure discussed in the main text of this paper (see Fig. \ref{fig:v_diff}). A passing substructure induces well localized velocity perturbations with a tell-tale double peak morphology.}
\label{fig:dvz}
\end{figure*}

The spatial extent of the substructure prevents the large velocity changes associated with $b\sim{}b_{90}$ in (\ref{eq:dvs}). We model the finite size of the substructure by rescaling $M_s$ in a continuous fashion with the impact parameter. Specifically, we make the following ansatz that mimics the radial scaling of a dark matter halo with a truncated NFW profile \citep{1996ApJ...462..563N, 1997ApJ...490..493N} and scale radius $b_s$.
\[
M_s = \begin{cases} \mu_0 & \text{if $b\gg{}b_s$}, \\
                              \mu_1 b & \text{if $b\sim{}b_s$}, \\
                              \mu_2 b^2 & \text{if $b\ll{}b_s$}.
       \end{cases} 
\]

Without loss of generality we can align the $x-y$-plane s.t. a given disk star lies at $\vec{r_*}=(r_{*,x}, 0, r_{*,z})$ and we neglect the height of the disk (i.e., $r_{*,z}=0$). It will prove useful to describe the Cartesian components of $\vec{v}_s$ using spherical coordinates, i.e., $v_{s,x}=v_s\sin\Theta{}\cos{}\Phi$, $v_{s,y}=v_s\sin\Theta{}\sin\Phi$, and $v_{s,z}=v_s\cos\Theta{}$. 
We now compute the maximal change in the $z$ velocity component of disk stars due to a perturbing substructure with a given velocity $\vec{v}_s$. First, we maximize $\vert\Delta{}v_{*,z}\vert \propto{} \frac{\mu_n}{v_s} \frac{\vert{}b_z\vert{}}{b^{n-2}}$ over the polar angle $\Phi$ and then over the distance $r_*$. For $n\in{}[0,2]$ and a given $r_*$
\[
\frac{b_z}{b^{n-2}} = r_*^{n-1} \cos{}\Theta{} \sin{}\Theta \cos{}\Phi{} \left[ 1 - (\sin\Theta{}\cos\Phi)^2 \right]^{\frac{n}{2}-1}
\]
is maximized or minimized for $\cos{}\Phi=\pm{}1$. For $n=0$ ($n=1$, $n=2$) the magnitude $\vert\Delta{}v_{*,z}\vert$ increases (remains constant, decreases) with decreasing $r_*$. Hence, the maximum and minimum of the vertical velocity perturbations of disk stars occur at a distance $r_*\sim{}b_s / \cos\Theta{}$ from the impact point of the substructure and lie along the projected path of the substructure. The typical spatial extent of these velocity peaks is half their separation. The peak velocity changes are
\begin{equation}
\max{}\vert{}\Delta{}v_{*,z}\vert = \frac{2 G M(<b_s)}{v_s b_s}\vert{}\sin\Theta\vert = \frac{2 G M(<b_s)}{v_s b_s}\sqrt{1 - \frac{v_{s,z}^2}{v_s^2}}.
\label{eq:dvz}
\end{equation}

The corresponding results for the $x$, and $y$ velocity components can be derived in a similar manner. The results are
\begin{flalign}
\max{}\vert{}\Delta{}v_{*,y}\vert &= \frac{2 G M(<b_s)}{v_s b_s}\sqrt{1 - \frac{v_{s,y}^2}{v_s^2}} \label{eq:dvy} \\
\max{}\vert{}\Delta{}v_{*,x}\vert &= \frac{2 G M(<b_s)}{v_s b_s} \label{eq:dvx}.
\end{flalign}
An accurate measurement of $\max{}\vert{}\Delta{}v_{*,x}\vert$, $\max{}\vert{}\Delta{}v_{*,y}\vert$, and $\max{}\vert{}\Delta{}v_{*,z}\vert$ allows to infer $\Theta$, $\Phi$, and the combination $M(<b_s)/(v_s b_s)$ using (\ref{eq:dvz}-\ref{eq:dvx}).

In Fig.~\ref{fig:dvz} we show the z-component of the stellar velocity perturbation as predicted by equation (\ref{eq:dvs}) for a spatially extended substructure with mass $M(<b_s)=1.1\times{}10^8$ $M_\odot$ and $b_s=1.3$ kpc. These values as well as the velocity angles $\Theta$ and $\Phi$ and the relative speed $v_s$ are chosen to mimic the low speed, vertical orbit of the substructure discussed in the main text of this paper. The velocity disturbances reach a magnitude of $\sim{}1.2$ km s$^{-1}$ and have a spatial extent of a few kpc. The differential rotation of the stellar disk results in winding of the velocity disturbances. Nonetheless, as the right panel in Fig.~\ref{fig:dvz} shows, a pronounced kinematic double peak structure is expected to remain visible for $\sim{}10^8$ yr (see also Fig.~\ref{fig:v_z}).

So far we used a coordinate frame in which the unperturbed disk star is at rest. We now switch to a coordinate system in which the Galactic Center is at rest. For simplicity we assume that the star moves in the disk ($x$-$y$) plane along the $y$ direction and the substructure moves in the $y$-$z$ plane. In this case the sine of the inclination angle $\theta$ ($\theta$ is defined as the angle between the orbital plane of the substructure and the plane of the disk of the MW) is also the sine of the angle between the velocity of the disk star $\vec{v}_*$ and the velocity of the substructure $\vec{v}_h$ and we obtain
\[
\max{}\vert{}\Delta{}v_{*,z}\vert = 2\frac{G M(<b_s)}{V b_s}  f(\theta,\phi),
\]
with $f(\theta,\phi) = \frac{ \vert{}\cos\phi - \sin\phi\,\cos\theta \vert{} }{ 1 - \sin(2\phi)\,\cos\theta }$, $V = \sqrt{v_*^2 + v_h^2}$, $\sin\phi=v_h/V$, $\cos\phi=v_*/V$, $\phi\in{}\left[0,\frac{\pi}{2}\right]$, and $\cos\theta\in{}\left[-1,1\right]$.

\begin{figure}
\begin{tabular}{c}
\includegraphics[width=80mm]{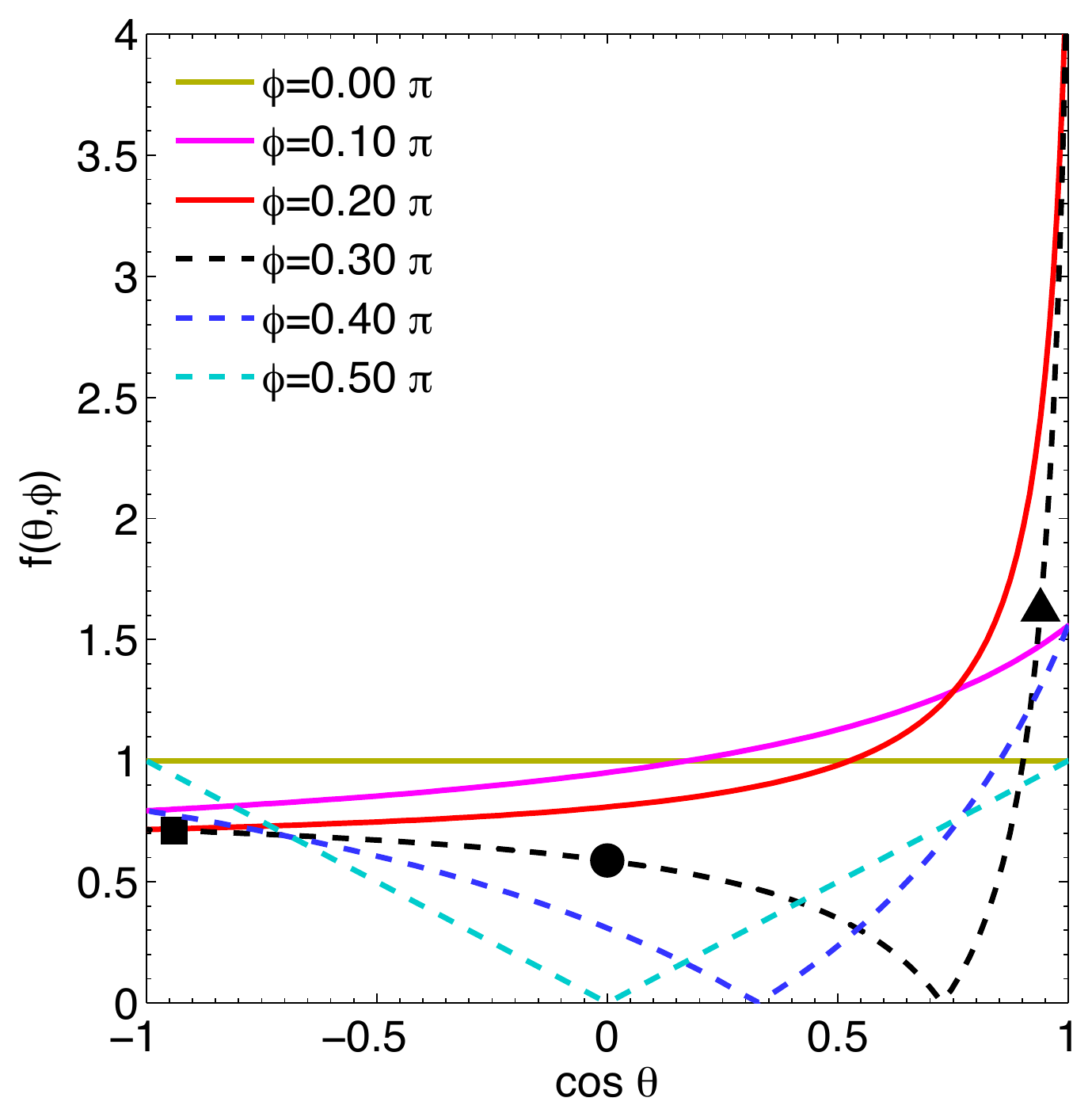}
\end{tabular}
\caption{Velocity factor $f(\theta,\phi)$ vs the cosine of $\theta$, the inclination angle of the orbit of the substructure. Each line corresponds to a particular choice of $\phi = \tan^{-1}(v_h/v_*)$. The solid (dashed) lines correspond to cases in which $v_h<v_*$ ($v_h>v_*$). The symbols show the expected velocity factors if we choose $\phi$ and $\theta$ values similar to the ones used in our numerical simulations. Specifically, we show the expected velocity factors for a retrograde ($\theta=160^\circ$, square), a vertical ($\theta=90^\circ$, circle), and a prograde ($\theta=20^\circ$, triangle) orbit of a substructure with $\phi=0.3\,\pi\approx{}0.94$. For given $v_h$ and $v_*$ with $v_h > v_*$ a vertical orbit leads to smaller vertical velocity changes than both retrograde or prograde orbits.}
\label{fig:velocityFactor}
\end{figure}

We plot the velocity factor $f(\theta,\phi)$ as a function of inclination angle in Fig.~\ref{fig:velocityFactor}. Clearly, the velocity factor is of order unity for many choices of $\theta$ and $\phi$. In addition, the figure demonstrates the following:

\begin{itemize}

\item If $\phi<\frac{\pi}{4}$, the speed of the substructure is smaller than the speed of the star, i.e., $v_h<v_*$. In this case the velocity factor decreases monotonically with inclination angle. The velocity factor can become large for $\theta\approx{}0$, i.e., if the substructure has a large velocity component along the motion of the star. In contrast, the velocity factor is reduced if the substructure has a large velocity component opposite to the motion of the star. We note that if $\phi\leq{}\frac{\pi}{4}$ the velocity factor never drops below $1/\sqrt{2}$.

\item If $\phi=\frac{\pi}{4}$, the velocity factor is $1/\sqrt{2}$ and independent of $\theta$. 

\item If $\phi>\frac{\pi}{4}$, the speed of the substructure is larger than the speed of the star,  i.e., $v_h>v_*$. In this case the velocity factor is non-monotonic and drops to zero for $\cos\theta=v_*/v_h$. 

\end{itemize}

These analytic predictions help us to understand why the change of $\Delta{}v_{*,z}$ in our numerical simulations is strongest for the prograde orbit, weakest for the vertical orbit of the substructure, and of intermediate strength for the retrograde orbit (see Fig. 3 of the paper). Here, $v_h\sim{}290$ km/s, $v_*\sim{}210$ km/s and, hence, $\phi=\tan{}^{-1}(290/210)\approx{}0.94>\frac{\pi}{4}$. Furthermore, $\theta=20^{\circ}$ for the prograde orbit, $\theta=90^\circ$ for the vertical orbit, $\theta=160^{\circ}$ for the retrograde orbit. Inserting $\phi$ and $\theta$ into the analytic expression for the velocity factor, we find $f\approx{}1.63$ for the prograde orbit, $f\approx{}0.59$ for the vertical orbit, and $f\approx{}0.71$ for the retrograde orbit. Hence, the prograde and retrograde orbits result in stronger changes of $\vert\Delta{}v_z\vert$ than a vertical orbit.

\section{Details of the simulation set-up}
\label{app:ModelSetup}

\paragraph*{Modeling of the MW:}
We use the tool GalactICS \citep{1995MNRAS.277.1341K, 2005ApJ...631..838W, 2008ApJ...679.1239W} to set up an approximately steady-state model of the MW, consisting of a stellar disk, a stellar bulge, and a dark matter halo. GalactICS simultaneously solves the collisionless Boltzmann equation and the Poisson equation of the total system to obtain close-to-equilibrium distribution functions for each of the specified galaxy components. It then samples these distribution functions to construct an N-body model of a galaxy. GalactICS requires various input parameters that determine the geometrical and kinematic properties of the galaxy.

Our MW model uses the parameters of the $Q=1.99$ \& $X=4.52$ model of \cite{2008ApJ...679.1239W}. Specifically, the stellar disk has mass $3.6\times{}10^{10}$ $M_\odot$, an exponential surface density profile with scale radius $2.84$ kpc, and a sech$^2$ density profile in the vertical direction with scale height $0.43$ kpc. The radial velocity dispersion $\sigma_R$ at the Galactic Center is $128.9$ km/s. The exponential scale length of $\sigma^2_R$ is $2.84$ kpc, i.e., the same as the scale length of the density of the stellar disk.

The stellar bulge has a density profile that yields the Sersic profile in projection \citep{1997A&A...321..111P} with Sersic index $n=1.28$, a mass of $1.0\times{}10^{10}$ $M_\odot$, a projected half mass radius of $0.556$ kpc and a velocity scale, see \cite{2008ApJ...679.1239W}, of 289.6 km/s. The dark matter halo has an NFW profile \citep{1996ApJ...462..563N, 1997ApJ...490..493N} of generalized form \citep{1990ApJ...356..359H,1996MNRAS.278..488Z} with a central cusp of $\gamma=0.977$, an outer slope of 3, a scale radius $a_h=9.39$ kpc and a velocity scale $\sigma_h=366.7$ km/s. We smoothly truncate the dark matter density beyond a radius of 200 kpc over a 20 kpc width. The mass of the halo within 200 kpc is $6.15\times{}10^{11}$ $M_\odot$.

As shown by \cite{2008ApJ...679.1239W} this galaxy model is in good agreement with observational data, see \cite{1998MNRAS.294..429D, 2002ApJ...574..740T} and references therein. For instance, the model reproduces the inner and outer rotation curves, the Oort constant, the vertical force in the solar neighborhood, the total mass at large radii, and the line-of-sight velocity dispersion of the bulge.

\begin{table}
\begin{center}
\begin{tabular}{l|l|l|l}
\tableline \tableline
label &$\theta$ & $\vec{x}=(x,y,z)$ & $\vec{v}=(v_x,v_y,v_z)$ \\ 
 & ($^\circ$) & (kpc) & (km/s) \\ \tableline
vertical &  90 &  (-12.91, 0, 22.66) & (123.3, 0, 34.22) \\
prograde &  20 &  (-12.91, -21.29, 7.75) & (123.3, -32.16, 11.71) \\
retrograde &  160 &  (-12.91, 21.29, 7.75) & (123.3, 32.16, 11.71) \\
vx(-400) & 90 & (-28.81, 0, -0.6045) & (279.5, 0, 61.59) \\
vx(+400) & 90 & (35.22, 0, 16.27) & (-226.8, 0, -144.1) \\
vx(+320) & 90 & (-26.54, 0, 13.49) & (297.3, 0, -57.93) \\
vx(-320) & 90 & (28.24, 0, 25.99) & (-155.3, 0, -233.6) \\
vx(+200) & 90 & (-17.86, 0, 23.88) & (245.1, 0, -156.8) \\
vx(-200) & 90 & (17.25, 0, 31.55) & (-52.13, 0, -274.2) \\
vz(450) & 90 & (-1.08, 0, 33.05) & (117.6, 0, -269.9) \\
\tableline \tableline
\end{tabular}
\caption{Setup of substructure -- MW interaction simulations. The first column provides the label of the respective simulation. The second column specifies the inclination angle of the orbit relative to the plane of the MW disk. The third and fourth columns list the initial position and velocity components of the substructure, respectively. The centers of the MW disk, bulge, and halo are initially at rest at the coordinate origin.}
\label{tab:ICs}
\end{center}
\end{table}

\begin{table*}
\begin{center}
\begin{tabular}{l|l|l|l|l|l|l|ll}
\tableline \tableline
label & speed & orbit type & $v_{h}$ & $(x, y)_0$ & $(v_{x},v_{y},v_{z})_0$ & $v_{s}$ & $(\Theta,\Phi)$ & expected $\max\vert\Delta{}v_z\vert$ \\ 
 & & & (km/s) & (kpc) & (km/s) & (km/s) & ($^\circ$) & (km/s) \\ \tableline
vertical & low & vertical & 288 & (11.45, 0) & (-4.6, 0, -288) & 356 & (144, -91) & 1.20 \\
prograde & low & incl. \& progr. & 292 & (10.97, -1.96) & (12.4, 270, -110)  & 129 & (148, 121) & 2.96 \\
retrograde & low & incl. \& retrogr. & 291 & (11.05, 1.68) & (8.3, -270, -108)  & 492 & (103, -93) & 1.45 \\
vx(+400) & high & vert. \& outw. & 446 & (7.1, 0) & (398, 0, -202)  & 493 & (114, -28) & 1.35 \\
vx(-400) & high & vert. \& inw. & 448 & (7.7, 0) & (-400, 0, -202)  & 495 & (114, -152) & 1.34 \\
vx(+320) & high & vert. \& outw. & 451 & (7.7, 0) & (319, 0, -319)  & 498 & (130, -33) & 1.12 \\
vx(-320) & high & vert. \& inw. & 448 & (8.5, 0) & (-317, 0, -317)  & 495 & (130, -146) & 1.13 \\
vx(+200) & high & vert. \& outw. & 445 & (7.8, 0) & (194, 0, -400)  & 492 & (144, -47) & 0.86 \\
vx(-200) & high & vert. \& inw. & 449 & (8.1, 0) & (-203, 0, -400)  & 495 & (144, -134) & 0.87 \\
vz(450) & high & vertical & 452 & (8.2, 0) & (0, 0, -452)  & 498 & (155, -90) & 0.62 \\
\tableline \tableline
\end{tabular}
\caption{Complete set of substructure -- MW interaction simulations. Labels for each simulation are given in the first column. The second and third columns highlight whether the substructure moves at low ($v_h\sim{}290$ km s$^{-1}$) or high ($v_h\sim{}450$ km s$^{-1}$) speed through the disk of the MW and provides a short description of the overall orbit type, respectively. The substructure speed in the galactocentric restframe is shown in the fourth column. Columns 5 and 6 provide the $x-y$ galactocentric coordinates and the velocity components of the density peak of the  substructure as it moves through the disk.  The centers of the MW disk, bulge, and halo are at rest at the coordinate origin. Column 7 estimates the relative speed between the substructure and a disk star near the point of impact. Column 8 provides the  polar angle $\Theta$ and the azimuthal angle $\Phi$ of the relative velocity between the substructure and the disk star, see Appendix \ref{app:VelChange}. The final column states the expected change in the vertical velocity according to equation (\ref{eq:dvz}) for a passing substructure with $M_s(<b_s)=1.1\times{}10^8$ $M_\odot$ and $b_s=1.3$ kpc. The passage of a $10^8$ $M_\odot$ perturber through the disk of the MW results in localized changes of $v_z$ of $0.6-3$ km s$^{-1}$ depending on the specific orbit of the substructure.}
\label{tab:SubstructureCollision}
\end{center}
\end{table*}

\paragraph*{Modeling of the substructure:}
We construct an N-body model of the orbiting dark matter substructure using a standard procedure widely used in the literature  \citep{1993ApJS...86..389H, 1999MNRAS.307..162S}. The substructure has a conventional NFW profile with concentration $c=17$ and a virial mass $M_h=1.1\times{}10^9$ $M_\odot$ within the radius $R_h=21.4$ kpc. The mass within the scale radius $R_h/c=1.3$ kpc is $1.1\times{}10^8$ $M_\odot$. We smoothly truncate the density outside $R_h$. The average density within $R_h$ corresponds to an overdensity of 200 times the critical density in the present universe. The virial velocity of the substructure is 15 km/s.

\paragraph*{Setup of the substructure -- MW interaction:}
We create the appropriate starting position and velocity of the substructure for the vertical case by running a lower resolution simulation of the inverted problem. Specifically, we chose a coordinate frame in which the MW model is centered on the coordinate origin and the angular momentum of the MW disk points in the $z$-direction. We then place the center of the substructure on the $x$-axis a certain distance (7-12 kpc) from the Galactic Center. We further give the substructure velocity components in the $x$ and $z$-directions. The subhalo speed is $\sim{}290$ km s$^{-1}$ ($\sim{}450$ km s$^{-1}$) in the low (high) speed set-up. We evolve this system forward past the point at which the substructure turns around and falls back towards the disk.We then record the position $\vec{x}$ and the velocity $-\vec{v}$ of the density peak of the substructure. Subsequently, we create the initial conditions for the actual substructure -- MW simulation by placing the center of the substructure at position $\vec{x}$ and by assigning the substructure the center-of-mass velocity $\vec{v}$. We create initial conditions for the inclined cases by rotating the orbital plane of the substructure around the $x$-axis.The specific initial positions and velocities of the substructure for each run can be found in Table~\ref{tab:ICs}.

\section{Simulations with additional orbital parameters}
\label{app:AddSims}
We provide information about the orbital parameters of each substructure -- MW collision simulations in Table~\ref{tab:SubstructureCollision}, including the speed of substructure at impact, the impact coordinates, and velocity components. We use the latter information to estimate the maximum change of the vertical velocity according to equation (\ref{eq:dvz}). We expect that the additional (high speed) encounters result in velocity changes of somewhat reduced amplitude compared with the low velocity collisions discussed in the main text. Furthermore, the amplitudes should depend on the inclination of the substructure orbit with respect to the plane of the MW disk. Specifically, orbits with higher in-plane velocity components should show a larger velocity impulse.

\begin{figure*}
\begin{tabular}{c}
\includegraphics[width=160mm]{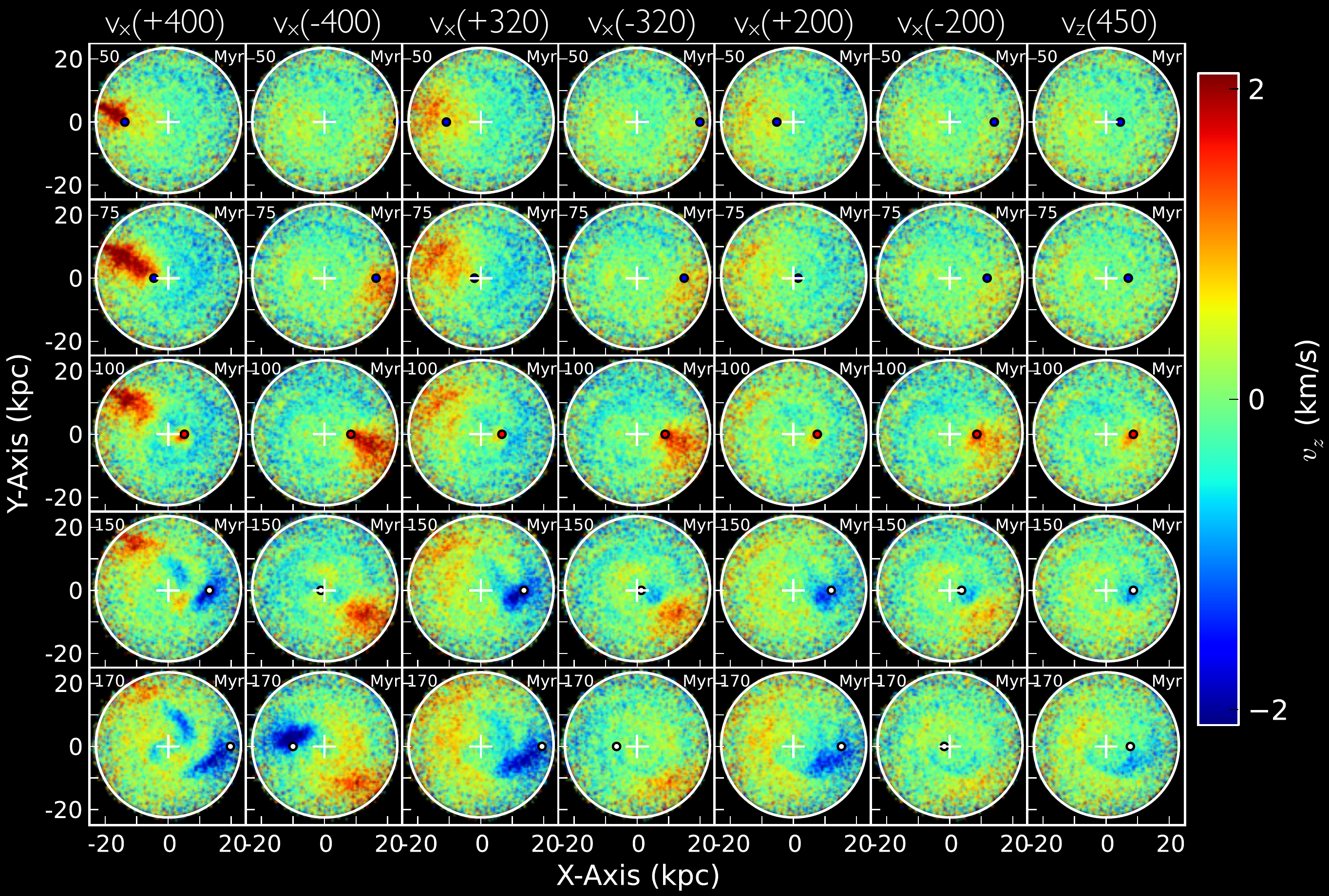}
\end{tabular}
\caption{Changes of the vertical velocity as measured in additional substructure -- MW disk encounters. Compared with the runs discussed in the main text the substructure speed is larger ($\sim{}450$ km s$^{-1}$) and most of the runs have a velocity component in the x-direction, see label at the top and Table~\ref{tab:SubstructureCollision}. Columns correspond to specific simulations and rows to time (see legend). The middle row corresponds to the time when the substructure passes through the MW disk. The velocity impulse imparted by the colliding substructure decreases from left to right as predicted by equation (\ref{eq:dvz}).}
\label{fig:MultiPlot}
\end{figure*}

These expectations are confirmed by Fig.~\ref{fig:MultiPlot}, which plots the vertical velocity changes for each of the additional (high speed) simulations at multiple epochs. The morphologies of the kinematic imprint show some variety, but are generally not too dissimilar from the results shown in Fig.~\ref{fig:v_z}. For instance, the $v_z(450)$, $v_x(+200)$, and $v_x(-200)$ runs look similar to the ``vertical'' case shown in the middle column of Fig.~\ref{fig:v_z}, while the $v_x(+400)$, and $v_x(-400)$ share similarities (e.g., the wedge-like early velocity peak and the tracing of the velocity trough by the substructure at late times) with the ``prograde'' case shown in the first column of Fig.~\ref{fig:v_z}.

\section{The angular and radial velocity of stars in the disk}

\begin{figure*}
\begin{tabular}{cc}
\includegraphics[width=80mm]{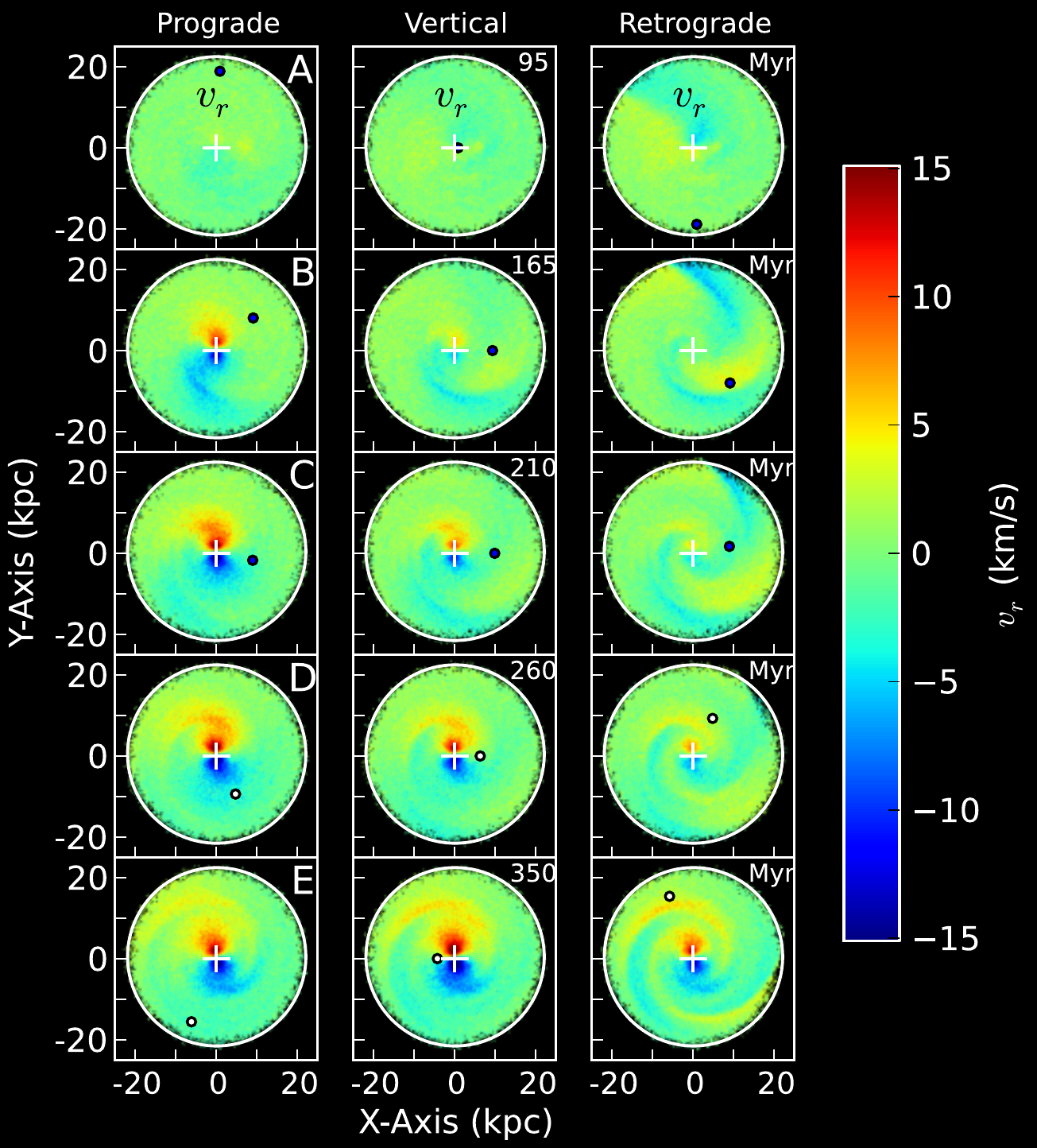} &
\includegraphics[width=80mm]{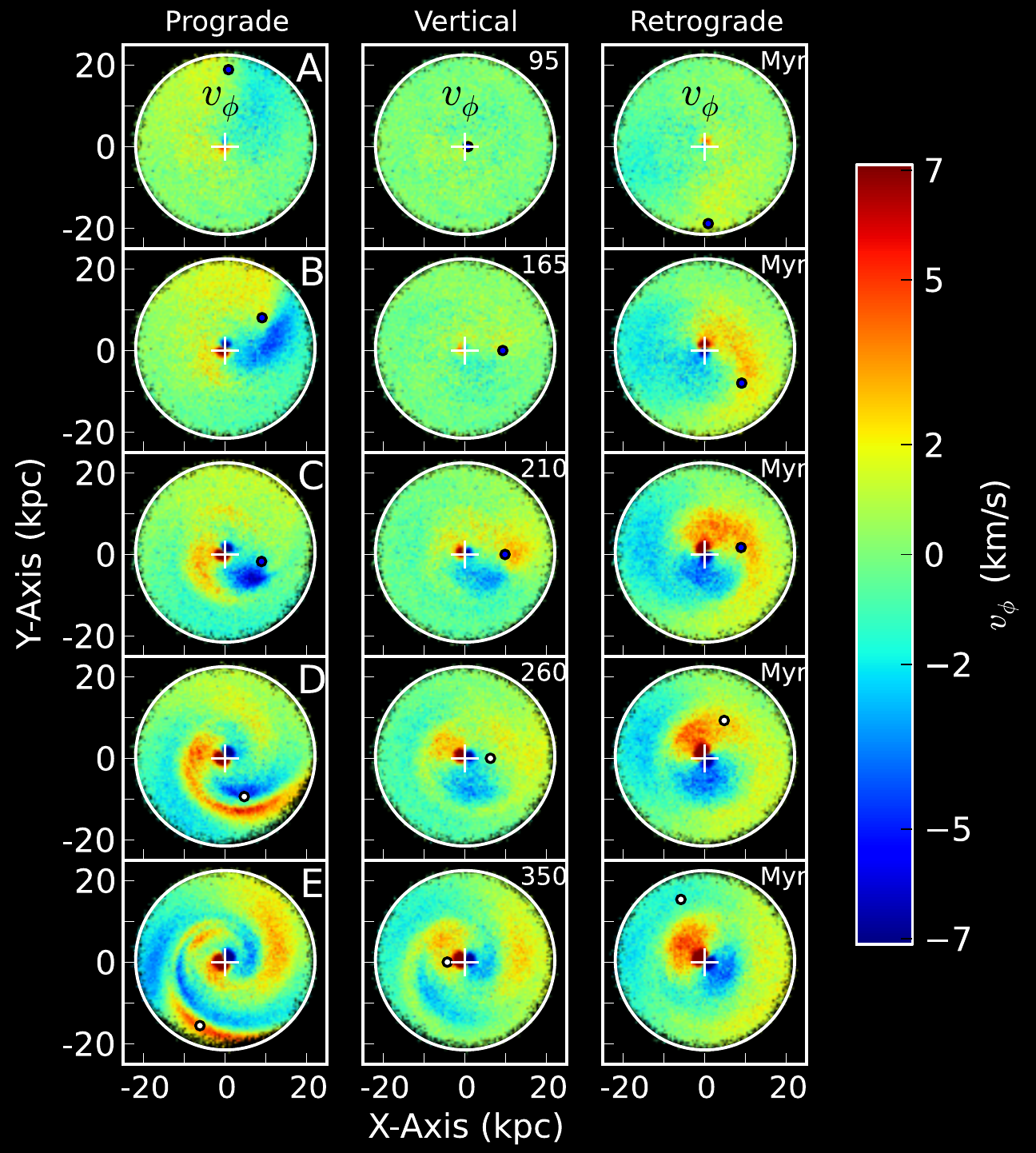}
\end{tabular}
\caption{Same as Fig.~\ref{fig:v_z} but for the changes of the  radial velocity $v_r$ (left panel) and azimuthal velocity $v_\phi$ (right panel) instead of the vertical velocity. We use a spatial binning of $500\times{}500$ pc$^2$. For the right panel we subtract the mean circular velocity from each particle prior to averaging.}
\label{fig:vr_500}
\end{figure*}

In the left panel of Fig.~\ref{fig:vr_500}, we plot the change of the mean velocity in the radial direction $r$. Similar to Fig.~\ref{fig:den} we notice a dipole-like structure and a wave, but now in velocity rather than in density. The velocity dipole is oriented perpendicular to the impact site of the substructure and has opposite polarity for the prograde and retrograde case.  At late times stars move in bulk away from the substructure in the prograde case and stars move toward the substructure in the retrograde case. We note that the density dipole described in the paper is rotated by $\sim{}90$ degrees with respect to the velocity dipole in Fig.~\ref{fig:vr_500}.   
In sum, perturbations in the mean motion of $v_r$ track the motion of density perturbation more so than the motion of the satellite. 

In the right panel of Fig.~\ref{fig:vr_500}, we plot the change of mean angular velocity  $v_\phi$. The disk rotates in a clockwise manner in this figure.
We  have subtracted the average rotation velocity of each star prior to binning. In panels A-D in the prograde case the bulk motion (positive/negative for a satellite above/below the disk) correlates with the position of the satellite. A similar result can be seen in panels B and C in the retrograde and vertical cases. In panel A of the middle column (vertical orbit) the substructure has yet to perturb the motion of the disk significantly. In the bottom row (140 Myr after the substructure -- disk collision), a velocity perturbation persists, but does not clearly track the motion of the disk.
\end{document}